\documentclass[amsmath,amssymb,11pt,superscriptaddress,reprint, preprintnumbers, notitlepage,aps,twocolumn,nofootinbib]{revtex4-1}
\pdfoutput=1 
\usepackage[utf8]{inputenc}
\usepackage[english]{babel}
\usepackage{amsmath}
\usepackage{graphicx}
\usepackage{dcolumn}
\usepackage{pbox}
\usepackage{amssymb}
\usepackage{epsfig}
\usepackage{slashed}
\usepackage{amssymb}
\usepackage{ mathrsfs }
\usepackage{color}
\usepackage[font=small]{caption}
\usepackage[font=small]{subcaption}
\usepackage{url}

\definecolor{MyLightBlue}{rgb}{0.22,0.51,0.9}
\definecolor{BrickRed}{rgb}{0.8, 0.25, 0.33}
\RequirePackage{hyperref}
\hypersetup{colorlinks, citecolor=blue,linkcolor=BrickRed, urlcolor=MyLightBlue}

\makeatletter
\renewcommand\@makecaption[2]{%
  \par
  \vskip\abovecaptionskip
  \begingroup
  
   \small\rmfamily
    \begingroup
     \samepage
     \flushing
     \let\footnote\@footnotemark@gobble
     \@make@capt@title{#1}{#2}\par
    \endgroup
  \endgroup
  \vskip\belowcaptionskip
}
\makeatother

\DeclareUnicodeCharacter{2212}{-}
\begin{document}
\title{\Large Non-Abelian Vector Dark Matter and Lepton $g-2$
}

\author{\bf Talal  Ahmed  Chowdhury}
\email[E-mail: ]{talal@du.ac.bd}
\affiliation{Department of Physics, University of Dhaka, P.O. Box 1000, Dhaka, Bangladesh}
\affiliation{The Abdus Salam International Centre for Theoretical Physics, Strada Costiera 11, I-34014, Trieste, Italy}

\author{\bf Shaikh Saad}
\email[E-mail: ]{shaikh.saad@unibas.ch}
\affiliation{Department of Physics, University of Basel, Klingelbergstrasse\ 82, CH-4056 Basel, Switzerland}

\begin{abstract}
The mystery of dark matter remains an unsettled problem of particle physics. On top of that,  experiments show a persistent contention of the muon anomalous magnetic moment (AMM) relative to the Standard Model (SM) prediction.  In this work, we consider the possibility of extending the SM with a non-Abelian gauge symmetry $SU(2)_X$, under which SM leptons transform non-trivially.  SM leptons receive corrections to their AMMs of right order via one-loop processes mediated by  beyond SM (BSM) fermions required to cancel anomalies, and BSM gauge bosons that play the role of dark matter. We show that simultaneous explanation of the the muon AMM along with reproducing correct relic abundance allows rather a narrow range of 0.5 - 2 TeV dark matter mass, consistent with current experimental constraints.  However, a concurrent description that also includes electron AMM is challenging in this set-up.        
\end{abstract}

\maketitle
\section{Introduction}
\vspace{-0.1in}
The existence of dark matter (DM) \cite{Oort,Zwicky:1933gu} is a one of the prevailing puzzles in particle physics. One of the most favored approaches to this problem exploits the fact that Weakly Interacting Massive Particles (WIMPs) in thermal equilibrium produce the dark matter relic abundance in the correct ballpark. The most successful theory in particle physics, the Standard Model (SM) is devoid of any DM candidate\footnote{Some possible dark matter candidates within the SM, although disfavored, have been explored in \cite{Farrar:2017eqq, Gross:2018ivp}.}. In this work, we explore the possibility of extending the SM with $SU(2)_X$ gauge symmetry and consider non-Abelian vector bosons to be the WIMPs.  In this framework, the SM leptons, both left-handed and right-handed ones transform non-trivially under  $SU(2)_X$, which plays, among others,  a significant role in  obtaining the correct relic abundance via dark matter annihilation into SM leptons. Since SM leptons are charged under $SU(2)_X$, new fermions must be added for gauge anomaly cancellation.

A model where only the SM left-handed leptons are charged under extended $SU(2)$ sector is proposed in Ref.  \cite{London:1986dk}.  The possibility of neutral vector boson  as dark matter candidate arising from this type of $SU(2)$ sector has been considered in Refs. \cite{DiazCruz:2010dc, Bhattacharya:2011tr,Ma:2012xj}\footnote{A case where dark matter emerges from extended $SU(2)\times U(1)$ sector, see for example Ref. \cite{Davoudiasl:2013jma}.}. In addition to left-handed ones, right-handed SM leptons can also transform non-trivially under the added $SU(2)$ group, as suggested in Ref. \cite{Fornal:2017owa}. In this latter set-up, neutral vector boson remains a possible dark matter candidate and such a scenario is studied  in  Ref.  \cite{Ma:2021roh}.

Besides dark matter, the SM is currently under scrutiny due to the precise measurement of the muon anomalous magnetic moment (AMM)  $a_\mu$, which is extremely sensitive to physics beyond the SM (BSM). There has been a longstanding tension between  the  theoretical  prediction  and  the  value measured  at  the BNL E821 experiment  \cite{Bennett:2006fi}. The recently announced result of FNAL  E989 experiment \cite{Abi:2021gix}, which has a smaller uncertainty, is fully compatible with  the previous best measurement. A combined result of these two experiments shows a remarkably large deviation of $4.2\sigma$ with respect to the SM prediction \cite{Aoyama:2020ynm}.  Various BSM scenarios are proposed to  explain the observed significant  departure, for a most recent review see Ref. \cite{Athron:2021iuf}.

In addition to the muon AMM, the electron AMM $a_e$ is also measured in the experiments with unprecedented level of accuracy. The recent improved measurement  \cite{Parker:2018vye}  of the fine-structure constant using Caesium atom shows a smaller value for $a_e$ with $2.4\sigma$ confidence level\footnote{A more recent measurement utilizing Rubidium atom \cite{Morel:2020dww}  with an accuracy of 81 parts per trillion on the other hand shows somewhat consistency with the SM value of $a_e$. Contrary to \cite{Parker:2018vye}, this new result  \cite{Morel:2020dww} finds $\Delta a_e$ to be positive ($+1.6\sigma$), indicating a $5.4\sigma$ discordance between these two experiments. However, the latest result of \cite{Morel:2020dww} is completely in disagreement with their  previous measurement for unknown reason. This is why, in this work we only focus on the result presented in \cite{Parker:2018vye}.} compared to theory value \cite{Aoyama:2017uqe}. The deviation $\Delta a_\ell= a^{exp}_\ell-a^{SM}_\ell$ is positive for the muon, whereas, it is negative for the electron. Moreover, the ratio $\Delta a_e/\Delta a_\mu$ is somewhat larger in magnitude than the naive lepton-mass-scaling $m^2_e/m^2_\mu$, which evidently makes it challenging to explain both these deviations concomitantly\footnote{ Motivated by these results there have been a number of proposals made in the literature to simultaneously explain the muon and the electron AMMs \cite{Giudice:2012ms, Davoudiasl:2018fbb,Crivellin:2018qmi,Liu:2018xkx,Dutta:2018fge, Han:2018znu, Crivellin:2019mvj,Endo:2019bcj, Abdullah:2019ofw, Bauer:2019gfk,Badziak:2019gaf,Hiller:2019mou,CarcamoHernandez:2019ydc,Cornella:2019uxs,Endo:2020mev,CarcamoHernandez:2020pxw,Haba:2020gkr, Bigaran:2020jil, Jana:2020pxx,Calibbi:2020emz,Chen:2020jvl,Yang:2020bmh,Hati:2020fzp,Dutta:2020scq,Botella:2020xzf,Chen:2020tfr, Dorsner:2020aaz, Arbelaez:2020rbq, Jana:2020joi,Chua:2020dya,Chun:2020uzw,Li:2020dbg,DelleRose:2020oaa,Kowalska:2020zve,Hernandez:2021tii,Bodas:2021fsy,Cao:2021lmj,Mondal:2021vou,CarcamoHernandez:2021iat,Han:2021gfu,Escribano:2021css,CarcamoHernandez:2021qhf,Chang:2021axw,Jueid:2021avn} in various BSM set-ups.}.

Since the DM candidate non-trivially interacts with the lepton sector, 
the framework we consider in this work, namely, the neutral $SU(2)_X$ extension of the SM, it is then  tempting to propose a combined explanation of all aforementioned puzzles. Hence the  philosophy of our work is to postulate that the new physics (NP) contributions to the muon and the electron AMMs  are intimately related to the origin of DM.  Assuming the neutral vector bosons to be the WIMPs, we show that  reproducing DM relic abundance in the correct ballpark along  with satisfying experimental observation of the muon AMM restricts the DM mass ($M_X$) and the new gauge coupling ($g_X$) within a narrow range that are $M_X\sim 0.5 - 2$ TeV and $g_X\sim  0.2 - 0.8$, respectively, for a specific region of the parameter space of the model considered in this work. When the electron AMM is added to the aforementioned list of observables, this model is  highly disfavored.  In finding the permitted parameter space, collider constraints, e.g., LEP and LHC bounds as well as electroweak (EW) precision data play crucial role in our analysis. 

The manuscript is organized as follows. In Sec. \ref{modelsec}, we discuss the specifics of the model, such as the gauge symmetry,  the particle content, the associated interactions, and symmetry breaking effects. In Secs. \ref{AMMsec} and \ref{EXPsec}, we present NP contributions to lepton AMM  and associated constraints from various experiments, respectively. Detailed explanations of DM physics is presented in Sec. \ref{DMsec}. In Sec. \ref{RESULTsec}, our main results are illustrated, and finally we conclude in Sec. \ref{CONsec}.

\section{Model}\label{modelsec}
\textbf{Gauge group and Fermion fields.}-- 
We consider a framework where the SM is supplemented by $SU(2)_X$ gauge group and consider the possibility that both the left-handed and right-handed leptons are charged under it  \cite{Fornal:2017owa,Ma:2021roh}. This requires additional fermions to cancel gauge anomalies and an anomaly free set  of fermions (per generation) is given below:
\begin{align}
&L_L=\begin{pmatrix} 
\nu_L&N_L \label{F1}\\
e_L&E_L
\end{pmatrix}
\sim (1,2,-\frac{1}{2},2),
\\&
\widetilde e_R=
\left( e_R\; E^{\prime}_R \right)\sim (1,1,-1,2), 
\\&
\widetilde\nu_R=
\left( \nu_R\; N^{\prime}_R \right)\sim (1,1,0,2),
\\&
\widetilde\ell_R= \begin{pmatrix} N_R\\ E_R  \end{pmatrix} \sim (1,2,-\frac{1}{2},1),
\\&
E^{\prime}_L\sim (1,1,-1,1),\;\; N^{\prime}_L\sim (1,1,0,1). \label{F2} 
\end{align}
In the above set, family index is suppressed, and quantum numbers of the fields under the complete gauge group $SU(3)_c\times SU(2)_L\times U(1)_Y \times SU(2)_X$ are presented. 

To generate masses of the BSM fermions, $SU(2)_X$ must be spontaneously broken. The simplest possibility is to consider a SM singlet, which transforms as a doublet of $SU(2)_X$ that we denote by $\phi=\left( \phi_1, \phi_2 \right)^T \sim (1,1,0,2)$.  The  $SU(2)_X$ gauge symmetry can be completely broken by the VEV $\langle \phi_1\rangle= v_X$, which subsequently generates vectorlike masses for the BSM fermions $E, E^{\prime}$ and $N, N^{\prime}$ (except the right-handed neutrinos $\nu_R$).

As usual, the SM symmetry is broken by the Higgs doublet, $H=\left( H^+, H^0 \right)^T\sim (1,2,1/2,1)$ that acquires the usual VEV $\langle H \rangle=v= 174$ GeV. All of the SM fermions, including neutrinos receive Dirac type masses as a result of EW breaking. Breaking of the  EW symmetry   allows a mixing between the $E$ and $E^{\prime}$ states (and similarly for $N$ and $N^{\prime}$ states), which turns out to be crucial to provide significant contribution to lepton AMMs to be discussed later in the text.  

The Yukawa part of the Lagrangian consists of the following terms
\begin{align}
-\mathcal{L}_Y&= y_e \overline{\widetilde e}_R   H^* L_L + y_\nu \overline{\widetilde \nu}_R H\epsilon L_L- y_0 \overline{\widetilde \ell}_R \phi  \epsilon L_L
\nonumber \\& -
y_E \overline E^{\prime}_L \phi \epsilon\widetilde e_R
-
y_N \overline N^{\prime}_L \phi \epsilon\widetilde \nu_R
\nonumber \\&
+\hat{y}_E \overline E^{\prime}_L H^* \widetilde \ell_R
+\hat{y}_N \overline N^{\prime}_L H \epsilon \widetilde \ell_R, \label{yukawa}
\end{align}
here $\epsilon_{21}=-\epsilon_{12}=1$. Inserting VEVs of the scalars, the mass matrices for $E, N$ can be written as
\begin{align}
-\mathcal{L}_Y&= \begin{pmatrix} 
\overline{E}_R&\overline E^{\prime}_R
\end{pmatrix}
\begin{pmatrix} 
y_0 v_X&\hat{y}_E^\dagger v\\
y_e v& y^\dagger_E v_X
\end{pmatrix}
\begin{pmatrix} 
E_L\\E^{\prime}_L
\end{pmatrix}
\nonumber \\ & +
\begin{pmatrix} 
\overline N_R&\overline N^{\prime}_R
\end{pmatrix}
\begin{pmatrix} 
y_0 v_X&\hat{y}_N^\dagger v\\
y_\nu v&y^\dagger_N v_X
\end{pmatrix}
\begin{pmatrix} 
N_L\\N^{\prime}_L
\end{pmatrix}. \label{EE}
\end{align}
Not only the $E$ and $N$ sectors are decoupled from each other but also 
the SM fermions do not mix with these new states. Besides, the Dirac 
masses of the charged leptons and neutrinos are given by $m_e= y_e v$ (just like the SM case) and $m_\nu=y_\nu v$, respectively.   For the simplicity of our work, we will ignore the intergenerational mixings, which however can be trivially included.  We diagonalize these two matrices by the following bi-unitary rotations
\begin{align}
\mathcal{M}_E=V^\dagger    \mathcal{M}_E^{diag} U,\;\;  \mathcal{M}_N=V^\dagger_N    \mathcal{M}_N^{diag} U_N. 
\end{align}
Correspondingly, the mass eigenstates $E^{(i)}$ with $i=1,2$, are connected by the flavor eigenstates as follows
\begin{align}
\begin{pmatrix}
E_L\\E^{\prime}_L
\end{pmatrix}    
=U^\dagger \begin{pmatrix}
E^{(1)}_L\\E^{(2)}_L
\end{pmatrix},
\;\; \begin{pmatrix}
E_R\\E^{\prime}_R
\end{pmatrix}    
=V^\dagger \begin{pmatrix}
E^{(1)}_R\\E^{(2)}_R
\end{pmatrix}, \label{rotation}
\end{align}
and similarly for states $N_{L,R}$, with $U\to U_N$ and $V\to V_N$.   To keep our analysis simple, we restrict ourselves to real Yukawa couplings.

\textbf{Scalar sector.}--   As aforementioned, the scalar sector of this theory is very simple and consists of the SM Higgs doublet $H$ and SM singlet $\phi$.  The complete scalar potential takes the form
\begin{align}
-\mathcal{L}\supset V &= \mu^2_H H^\dagger H + \frac{1}{2} \lambda_H \left( H^\dagger H \right)^2 +  \mu^2_\phi \phi^\dagger \phi 
\nonumber \\&
+ \frac{1}{2} \lambda_\phi \left( \phi^\dagger \phi \right)^2   
+\lambda_m \left( H^\dagger H \right) \left( \phi^\dagger \phi \right). \label{scalar}
\end{align}
Owing to symmetry breaking, three real degrees of freedom from each of these fields are eaten up by the corresponding gauge bosons, leaving in total two real scalar degrees of freedom. The mass-squared matrix in a basis of $\left( \sqrt{2} Re[\phi_1]\; \sqrt{2} Re[H^0]\right)$ is given by
\begin{align}
\begin{pmatrix}
2\lambda_\phi v^2_X&2\lambda_m v v_X
\\
2\lambda_m v v_X&2\lambda_H v^2
\end{pmatrix}.    
\end{align}
Diagonalization of this matrix leads to two mass eigensates defined as follows
\begin{align}
&\begin{pmatrix}
\phi_X\\h
\end{pmatrix}   
=\begin{pmatrix}
c_\theta&s_\theta\\
-s_\theta&c_\theta 
\end{pmatrix}
\begin{pmatrix}
\sqrt{2} Re[\phi_1]\\ \sqrt{2} Re[H^0]
\end{pmatrix},
\\
&\tan 2\theta= \frac{2\lambda_m}{\lambda_\phi/r_v-\lambda_H\; r_v},\,\,\mathrm{where}\,\,r_v=v/v_X.
\end{align}
The mass eigenvalues of $\phi_{X}$ and $h$ are given as
\begin{equation}
    m^2_{\phi_{X},h}=\lambda_H v^2 +\lambda_\phi v^2_X \pm \bigg\{
\left(\lambda_\phi v^2_X - \lambda_H v^2 \right)^2 + 4 \lambda^2_m v^2v^2_X
\bigg\}^{1/2}.
\label{masseg}
\end{equation}

As the mass of the Standard Higgs boson is experimentally fixed to be $m_{h}=125.1$ GeV, we determine the scalar couplings, $\lambda_{H,\phi,m}$ in terms of Higgs mass, $m_{h}$, BSM neutral Higgs mass, $m_{\phi_{X}}$, and the corresponding mixing angle, $\theta$ as follows,
\begin{align}
    \lambda_{H}&=\frac{m_{h}^{2}+m_{\phi_{X}}^{2}-(m_{\phi_{X}}^{2}-m_{h}^{2})\cos\theta}{4 v^{2}},\label{lambda1}\\
    \lambda_{\phi}&=\frac{m_{h}^{2}+m_{\phi_{X}}^{2}+(m_{\phi_{X}}^{2}-m_{h}^{2})\cos\theta}{4 v_{X}^{2}},\label{lambda2}\\
    \lambda_{m}&=\frac{(m_{\phi_{X}}^{2}-m_{h}^{2})\sin 2\theta}{4 v v_{X}}\label{lambda3}.
\end{align}

\textbf{Gauge interactions.}--  Like the $SU(2)_L$ part of the SM, the added $SU(2)_X$ gauge factor comes with three  vector bosons. Note however that each of them are electromagnetically neutral. We denote these gauge bosons as $X= \left(X_1-i\;X_2\right)/\sqrt{2}$, $X^\dagger= \left(X_1+i\;X_2\right)/\sqrt{2}$, and $Z^{\prime}=X_3$.  Interactions of these newly introduced gauge bosons are given by
\begin{align}
&\mathcal{L}_G \supset
\frac{g_X}{2} Z^{\prime}_\mu \bigg\{  
 \overline{e}\gamma^\mu e +  \overline{\nu}\gamma^\mu \nu - \overline E_L\gamma^\mu E_L -  \overline E^{\prime}_R\gamma^\mu E^{\prime}_R
 \nonumber \\&
 - \overline N_L\gamma^\mu N_L -  \overline N^{\prime}_R \gamma^\mu N^{\prime}_R
\bigg\} 
+
\frac{g_X}{\sqrt{2}} X_\mu \bigg\{   \overline \nu_L \gamma^\mu N_L + \overline e_L \gamma^\mu E_L
\nonumber \\&
+ \overline \nu_R \gamma^\mu N^{\prime}_R + \overline e_R \gamma^\mu E^{\prime}_R   + h.c. \bigg\}. \label{gaugeX}
\end{align}
Here $g_X$ is the gauge coupling associated to $SU(2)_X$ group. When $\phi$ develops VEV, all the gauge bosons acquire degenerate mass given by $M_X= \frac{1}{2} g^2_X v_X^2$.

A spectacular feature of this model is that when  the $SU(2)_X$ local symmetry is spontaneously broken, a residual global $U(1)$ symmetry emerges from it \cite{Ma:2021roh}. Looking at all the interactions of this theory, viz, Eqs. \eqref{yukawa}, \eqref{scalar}, and \eqref{gaugeX},    it is clear that all the SM fermions along with right-handed neutrino $\nu_R$, the BSM neutral Higgs $\phi_X$, and $Z^{\prime}$ are neutral under this $U(1)$,  whereas $E, E^{\prime}, N, N^{\prime},$ and the vector boson $X^\dagger$, each carry one unit of dark charge $Q_X=+1$.   We identify this as the dark    $U(1)_X$ symmetry, which is responsible for stabilizing the dark matter in our set-up.  In this work, we consider a scenario where the gauge boson $X$ is the dark matter candidate, hence must be the lightest among $\{E, E^{\prime}, N, N^{\prime}, X\}$.

It is to be notated that in this theory the SM gauge bosons receive additional interactions that can potentially affect the EW precision data. Consequences of these interactions are discussed in Sec. \ref{EXPsec}.

\section{Lepton AMM}\label{AMMsec}
First we briefly summarize the current experimental status of the lepton anomalous magnetic moments defined as $a_\ell=(g_\ell-2)/2$.  Since AMMs for the muon and the electron are very precisely measured quantities, they provide excellent tests of physics beyond  the  SM.

The previous measurement ($a_\mu= 116592089(63)\times 10^{-11}$) of $a_\mu$ from BNL \cite{Bennett:2006fi} about two decades ago showed a significant deviation from the SM prediction ($a_\mu= 116591810(43)\times 10^{-11}$) that corresponds to a positive $3.7\sigma$ discrepancy. This longstanding tension just recently has been confirmed by the FNAL result \cite{Abi:2021gix} ($a_\mu= 16592040(54)\times 10^{-11}$), which has smaller uncertainty. Their respective deviations relative to the SM value correspond to
\begin{align}
&\Delta a_\mu^{BNL} = (2.79\pm 0.76)\times 10^{-9}, \\
&\Delta a_\mu^{FNAL} = (2.30\pm 0.69)\times 10^{-9}.
\end{align}
Combinedly  these two results point towards a large $4.2\sigma$ tension with SM value:
\begin{align}
&\Delta a_\mu^{comb} = (2.51\pm 0.59)\times 10^{-9}.
\end{align}

As for the electron, a recent measurement performed at the Berkeley National Laboratory \cite{Parker:2018vye} yields a smaller $a_e$ than the SM prediction. Their result shows a  deviation given by
\begin{align}
&\Delta a_e = (-8.8\pm 3.6)\times 10^{-13},
\end{align}
which corresponds to $2.4\sigma$ disagreement from SM value.

Note that the quantity $a_\ell$ is flavour conserving, CP-conserving,  chirality flipping, and must be loop induced. In the SM and in many BSM extensions, this chiral symmetry is broken only by the non-vanishing mass term $m_\ell$ for the corresponding lepton. Consequently, a relation of the form  $a_\ell\propto m_\ell^2$ holds, which makes BSM contributions to be small. It is somewhat challenging to find a common BSM origin to  resolve both the muon and the electron AMMs, not only because the magnitude of their relative deviations is larger than the naive mass scaling $m^2_e/m^2_\mu$, but also due to  their opposite signs.  To provide large corrections to both the muon and the electron AMMs, as suggested by experimental results,  additional  sources  of chiral  symmetry breaking  of the muon and the electron are required.  

\begin{figure}[th!]
\includegraphics[width=0.4\textwidth]{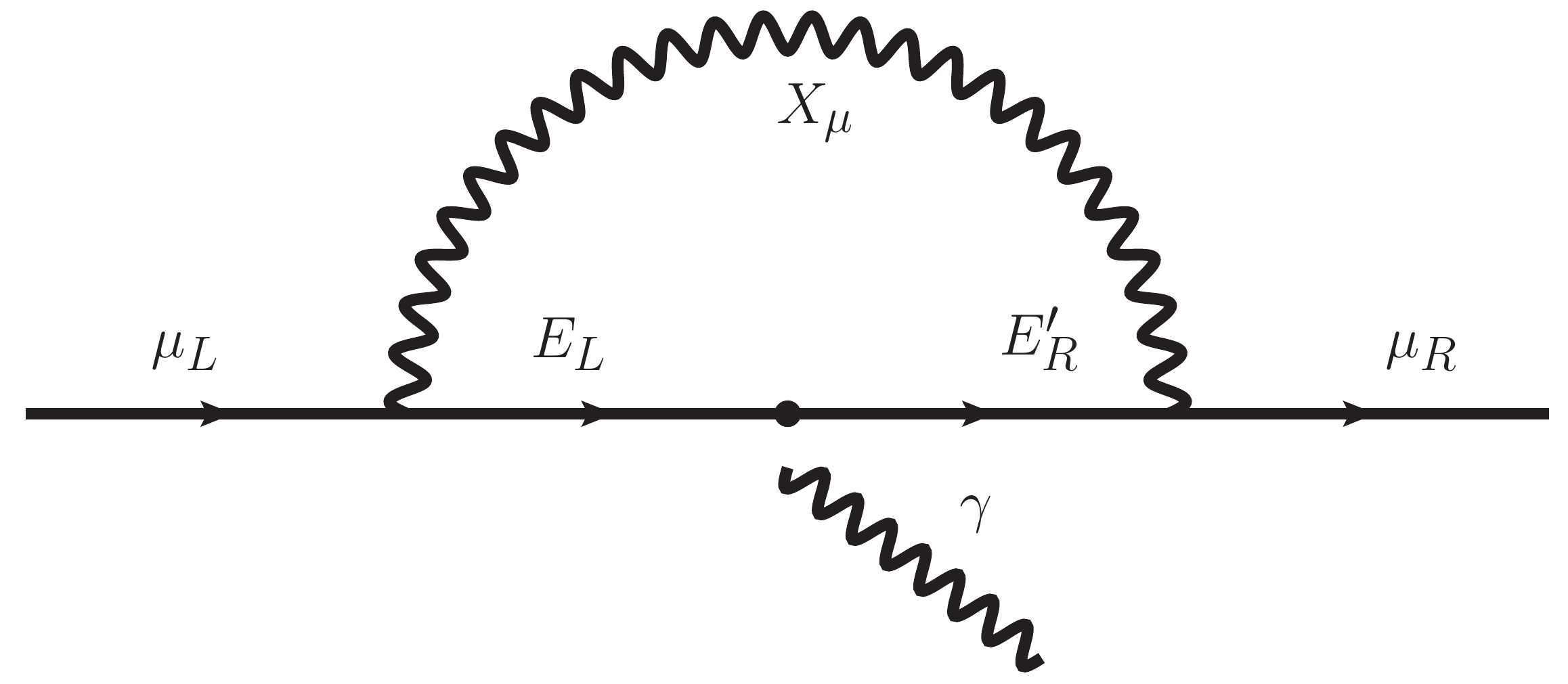}
\caption{Leading order contribution to muon $a_\mu$, here $E, E^{\prime}$ refer to muon-like heavy lepton. A similar diagram for the electron can be drawn with electron-like heavy leptons running inside the loop.} \label{muon}
\end{figure}

The model presented in this work, such a chirality flipping contribution appears at the one-loop order via the dark matter exchange as shown in Fig. \ref{muon}.  It is crucial to realize that even though breaking of $SU(2)_X$ generates vectorlike masses for $E, E^{\prime}$, they are  allowed to mix only after EW symmetry is broken, see Eq. \eqref{EE}. New physics contribution to lepton AMMs of Fig. \ref{muon} vanishes in the SM unbroken phase.

Now, utilizing the rotations of the fermions fields defined in Eq. \eqref{rotation}, and gauge interactions of Eq. \eqref{gaugeX}, the relevant dark matter coupling for $a_\ell$ to fermions in the mass basis  can be written as 
\begin{align}
&\mathcal{L}\supset X_\mu \left(\Gamma^L_{\ell,i} \overline{E}_L^{i}\gamma^\mu \ell_L + \Gamma^R_{\ell,j} \overline{E}_R^{j}\gamma^\mu \ell_R  \right), 
\\&
\Gamma^L_{\ell,k}= \frac{g_X}{\sqrt{2}} U^{\ell}_{k1},\; \; \Gamma^R_{\ell,k}= \frac{g_X}{\sqrt{2}} V^{\ell}_{k2},
\end{align}
where sum over repeated indices is understood. For concreteness, here we have put a superscript of $\ell$ on $U, V$ to distinguish rotation matrices for different flavors involved.  With all these, we derive the BSM contribution to lepton AMM to be \cite{Leveille:1977rc}  
\begin{align}
a^{BSM}_\ell= -\frac{m_\ell}{4\pi^2m^2_X} &\bigg\{ Re\left[
\Gamma^{L*}_k\Gamma^R_k\right] M^{(k)}_EF[x_k] 
\nonumber \\&
+ m_\ell \left(\left| \Gamma^L_k \right|^2+ \left| \Gamma^R_k \right|^2 \right) G[x_k]
\bigg\},
\end{align}
and the loop functions are given by ($\sqrt{x_k}=M^{(k)}_E/M_X$)
\begin{align}
&F[x]=\frac{4-3x-x^3+6x\ln[x]}{4(x-1)^3},
\\
&G[x]=\frac{8-38x+39x^2-14x^3+5x^4-18x^2\ln[x]}{24(x-1)^4}.
\end{align}

\begin{figure}[th!]
\includegraphics[width=0.35\textwidth]{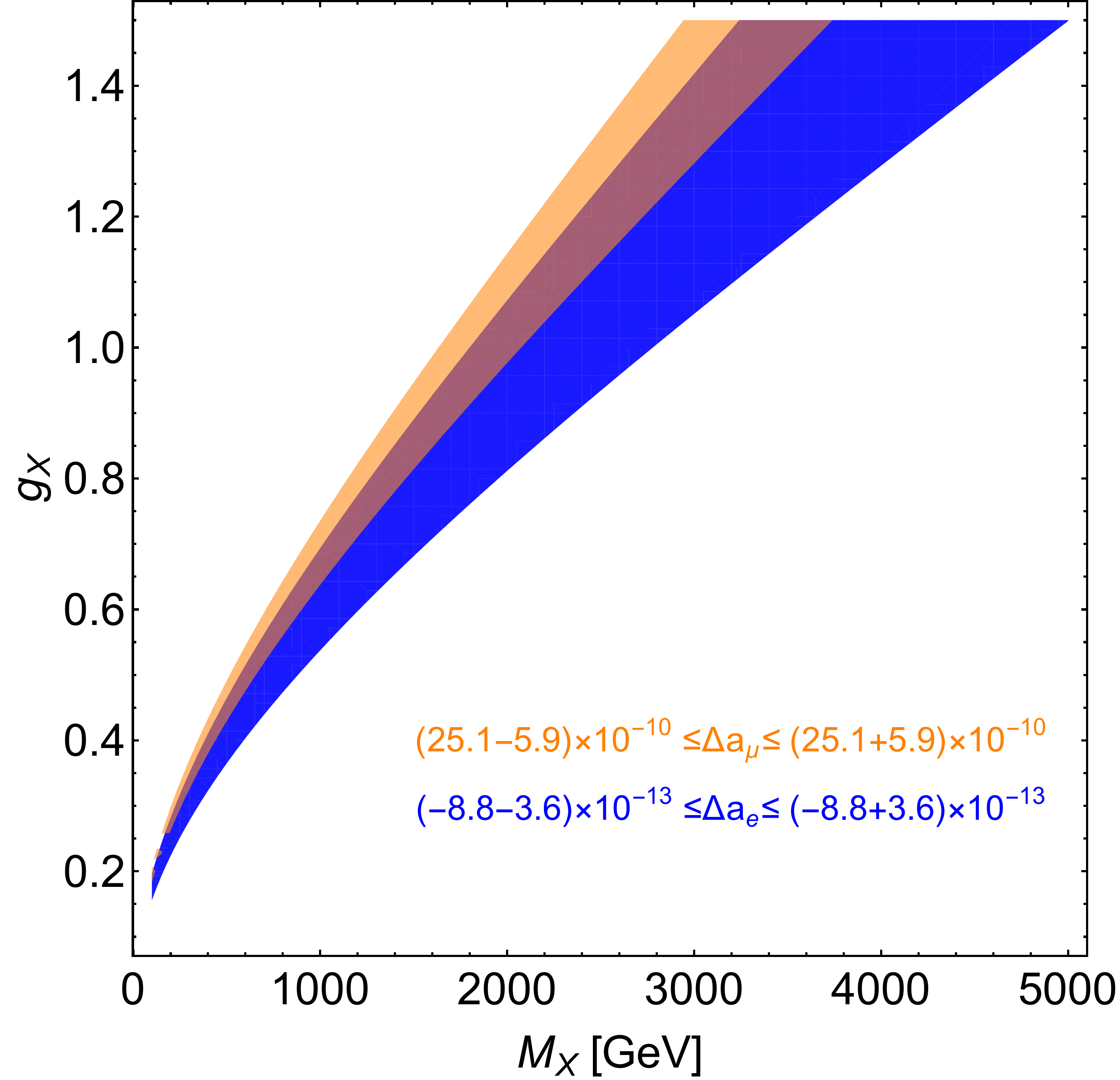}
\caption{BSM corrections to lepton AMMs arising from Fig. \ref{muon}. The orange (blue) band corresponds to $g-2$ for the muon (electron) within its $1\sigma$ experimental value. The overlapping parameter space, i.e, the brown band is where both are satisfied. For details, see text.} \label{gM2}
\end{figure}

From this, one sees that  the first term dominates due to its chiral enhanced effect, and rest of the terms can be ignored. For numerical analysis, we however, use the full expression.  For later convenience, in Fig. \ref{gM2}, we demonstrate the dependence of these corrections to lepton AMMs, on the two most crucial parameters of the theory, namely, $g_X$ and $M_X$. The orange (blue) band corresponds to $\Delta a$ for the muon (electron) within its $1\sigma$ experimental value. The overlapping parameter space (brown band) shows the required values in the $M_X - g_X$ plane to simultaneously incorporate $\Delta a_\ell$.  In this plot, we have fixed the relevant Yukawa couplings to be  $y^{e,\mu}_0= y^{e,\mu}_E= 1.5= -3 \hat y^\mu_{E}= 30 \hat y^e_{E}$.

\section{Experimental constraints}\label{EXPsec}
Here we summarize relevant experimental constraints of our model.

\begin{figure}[th!]
\includegraphics[width=0.35\textwidth]{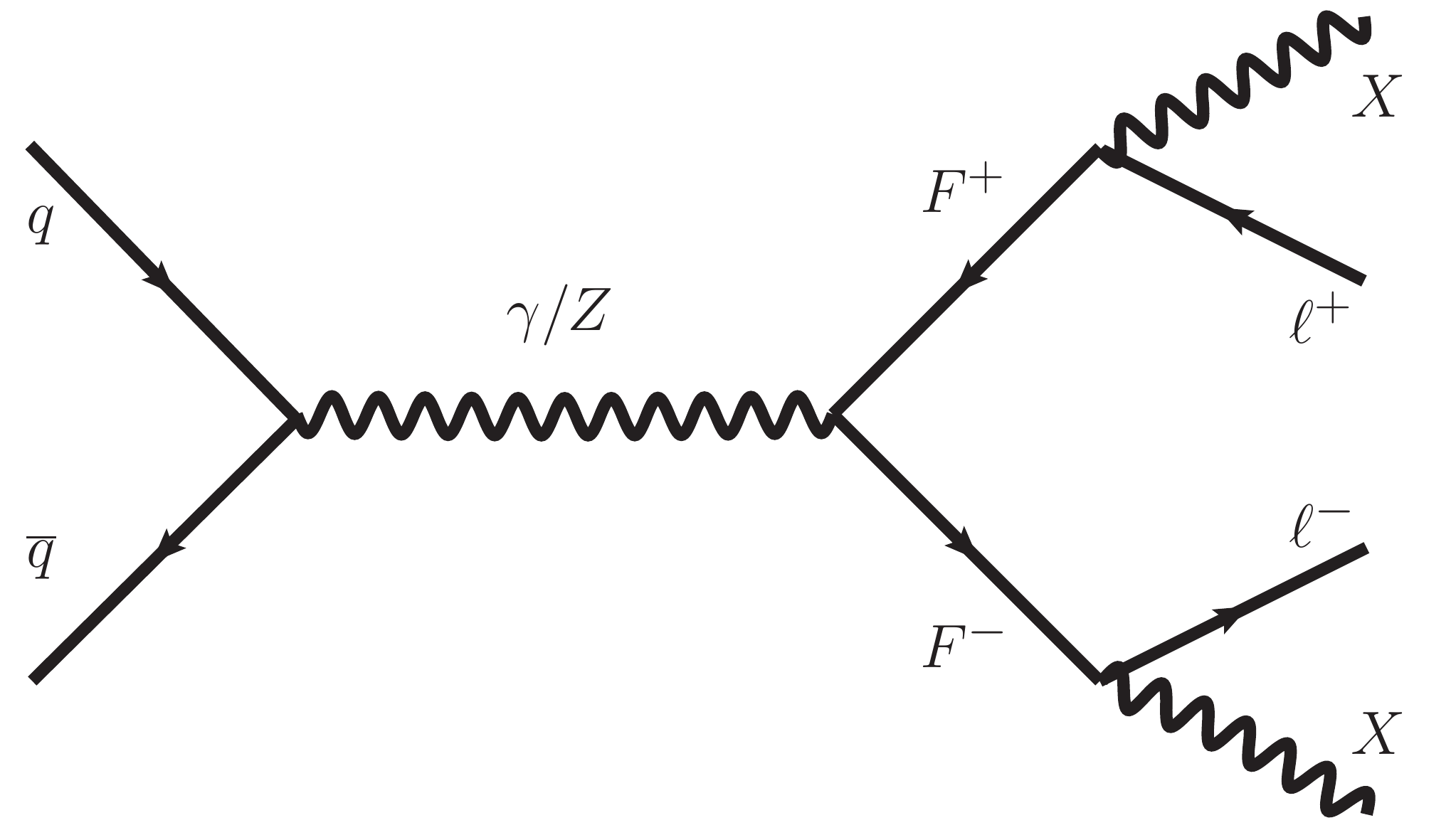}
\caption{Representative Feynman diagram leading to possible $pp\to \ell^-\ell^+ + \slashed{E}_T$ at the LHC.} \label{collider}
\end{figure}

\textbf{LHC constraints.}-- 
The large hadron collider (LHC) is searching for charged fermions beyond the SM. For each flavor, we have two types of singly charged fermions that we commonly denote as $F^{\pm}= E^{\pm}, E^{\prime \pm}$. Even though our BSM fermion $F^{\pm}$ has no interactions with quarks, they can still be pair produced at LHC
via s-channel $\gamma/Z$ exchange as displayed in Fig. \ref{collider}. This Feynman diagram shows that once produced, each $F^{\pm}$ will decay into a dark matter (lighter than $F^{\pm}$ in our scenario) and a SM charged lepton that gives rise to $pp\to \ell^-\ell^+ + \slashed{E}_T$. Processes of this type are constrained by LHC due to the standard slepton searches \cite{Aad:2014yka,Sirunyan:2018nwe,Sirunyan:2018vig}. Assuming the existence of both the left-handed and right-handed partners, as in our case, LHC puts a lower limit of $450$ GeV for their masses \cite{Sirunyan:2018nwe}. 

\textbf{LEP constraints.}--
In addition to dark vector boson, 
since our model also contains a $Z^{\prime}$ that does not carry any dark charge, it directly decays to SM leptons as depicted in Eq. \eqref{gaugeX} (the first term). Processes like this are highly constrained by LEP experiment \cite{LEP:2003aa}. In fact there are two types of $Z^{\prime}$ searches, indirect and direct, and the former seems to provide stronger bound in our scenario. Direct bound is applicable for $Z^{\prime}$ mass below the center of mass scale of LEP-II that constraints $m_{Z^{\prime}}\leq 209$ GeV.  On the other hand, indirect bound  arises from  four fermi contact interaction leading to $e^+e^-\to f\overline{f}$  originating from integrating out $Z^{\prime}$ ($f$ is any SM fermion). The strongest bound comes from $e^+e^-\to \ell^+\ell^-$ final states, which for vectorial coupling corresponds to $\Lambda^{\ell^+\ell^-}_{VV}=24.6$ TeV \cite{ALEPH:2013dgf}.  Following the analysis performed in Ref. \cite{Carena:2004xs}, we find\footnote{Related future bounds from ILC can be found for example in Ref. \cite{Das:2021esm}.} the most stringent constraint from LEP-II that translates into $M_X/(g_X/2)> 6.94$ TeV ($\equiv \Lambda_{LEP}$) for our case.

\textbf{EW precision constraints.}-- 
As already pointed out, the SM gauge bosons have additional interactions in this model that alter the vacuum polarisation and lead to corrections to oblique parameters. In our set-up, mixing between doublets and singlets  play the vital role in explaining muon and electron AMMs, which subsequently contribute to these precision electroweak observables. We find that the strongest such constraints originate from $T$-parameter \cite{Peskin:1990zt} within this framework, which we take into account in our calculation.

The expression for the $T$-parameter from BSM fermions has the form \cite{Lavoura:1992np,Chen:2017hak}
\begin{align}
&\Delta T= \frac{1}{4\pi s^2_{2W}}\sum_{i,j} \bigg\{
\left( \left|A^L_{ij}  \right|^2 +  \left|A^R_{ij}  \right|^2 \right) F_+(w_i,w_j) 
\nonumber \\ &
+ 2 Re\left(A^L_{ij} A^{R*}_{ij} F_{-}(w_i,w_j) \right) -\frac{1}{2}\left( \left|B^L_{ij}  \right|^2 +  \left|B^R_{ij}  \right|^2 \right) 
 \\ &\nonumber 
 \times F_+(w_i,w_j) 
-  Re\left(B^L_{ij} B^{R*}_{ij} F_{-}(w_i,w_j) \right) 
\bigg\}, 
\end{align}
here $A\;g/\sqrt{2}$ and $B\;g/(2c_W)$ are the couplings of the $W^+$ and $Z$ bosons, respectively, after going to the mass basis of the fermions following Eqs.\ \eqref{rotation}. 
And the associated loop functions take the form
\begin{align}
&F_+(w_i,w_j)= w_1+w_2-\frac{2w_1w_2}{w_1-w_2}\ln\left[\frac{w_1}{w_2}\right], \\
&F_-(w_i,w_j)=2\sqrt{w_1w_2}\left(\frac{w_1+w_2}{w_1-w_2}\ln\left[\frac{w_1}{w_2}\right]-2\right).  
\end{align}
We impose the experimental $1\sigma$ bound on this parameter   $\Delta T=0.05\pm 0.06$ \cite{Zyla:2020zbs} in our numerical study.

\section{Dark Matter Relic Density and Direct Detection}\label{DMsec}
\vspace{-0.1in}
\textbf{Dark Matter Parameter Space.}--
As already mentioned in section \ref{modelsec}, the single-charged BSM fermions, $E$, $E'$ and the neutral fermions, $N$, $N'$, and $X^{\dagger}$ carry the conserved dark charge, $Q_{X}=1$, and compose the dark sector of this model. We consider $X$ to be the vector dark matter candidate\footnote{The vector dark matter can arise also from $U(1)$ extension of the SM, see for example \cite{Farzan:2012hh}.} in this work, and to avoid its decay into the BSM fermions and charged leptons, $l$ and neutrinos $\nu_{l}$, $X\rightarrow \overline{E}\,l,\,\overline{E'}\,l,\, \overline{N}\,\nu_{l},\,\overline{N'}\,\nu_{l}$, the mass of $X$ is set to $M_X<m_{E,E'},m_{N,N'}$. Before describing its relic abundance, let us delineate the relevant parameter space for the DM set by the Direct Detection experiments.

\textbf{Dark Matter Direct Detection.}-- At tree-level, the spin-independent DM-nucleon cross-section of $X$ is mediated by the SM Higgs exchange, and given as
\begin{equation}
    \sigma_{\mathrm{SI}}=\frac{1}{4\pi}\frac{|F_{nX}|^{2}\mu_{r}^2}{M_X^{2}}
    \label{dmdirect1}
\end{equation}
where, the effective coupling between $X$ and the nucleon, $n$ is determined as, $F_{nX}=\frac{g_{X}M_X\sin\theta}{m_{h}^{2}}\frac{f_{n}m_{n}}{v}$ following the prescription of \cite{Hisano:2010yh} and the reduced mass is $\mu_{r}=\frac{M_Xm_{n}}{M_X+m_{n}}$. Moreover, $m_{n}=0.938$ GeV is the nucleon mass and $f_{n}$ parametrizes the effective coupling between the Higgs boson and the nucleon, and is given by $f_{n}=0.308$ \cite{Hoferichter:2017olk}.

Moreover, the scalar couplings, $\lambda_{H,\phi,m}$ are determined in terms of the parameters $\{m_{h}, m_{\phi_X}, g_{X}, M_X,v, \theta\}$ using Eq. \ref{lambda1}, \ref{lambda2} and \ref{lambda3}, and we constrain them within the range, $0\leq \lambda_{H,\phi,m}\leq 1$ to ensure their perturbativity at larger energy scale. Combining these constraints with the limit on the spin-independent DM direct detection from XENON1T \cite{XENON:2018voc}\footnote{After the submission of this work, the PandaX-4T collaboration has presented a new limit on the spin-independent DM-nucleon interactions \cite{PandaX:2021osp} which can be relevant for such study.}, we determine the allowed region of $m_{\phi_X}-\theta$ for a specific value of $(M_X, g_{X})$.
\begin{widetext}
\begin{figure*}[th!]
    \center{\includegraphics[width=0.33\textwidth]{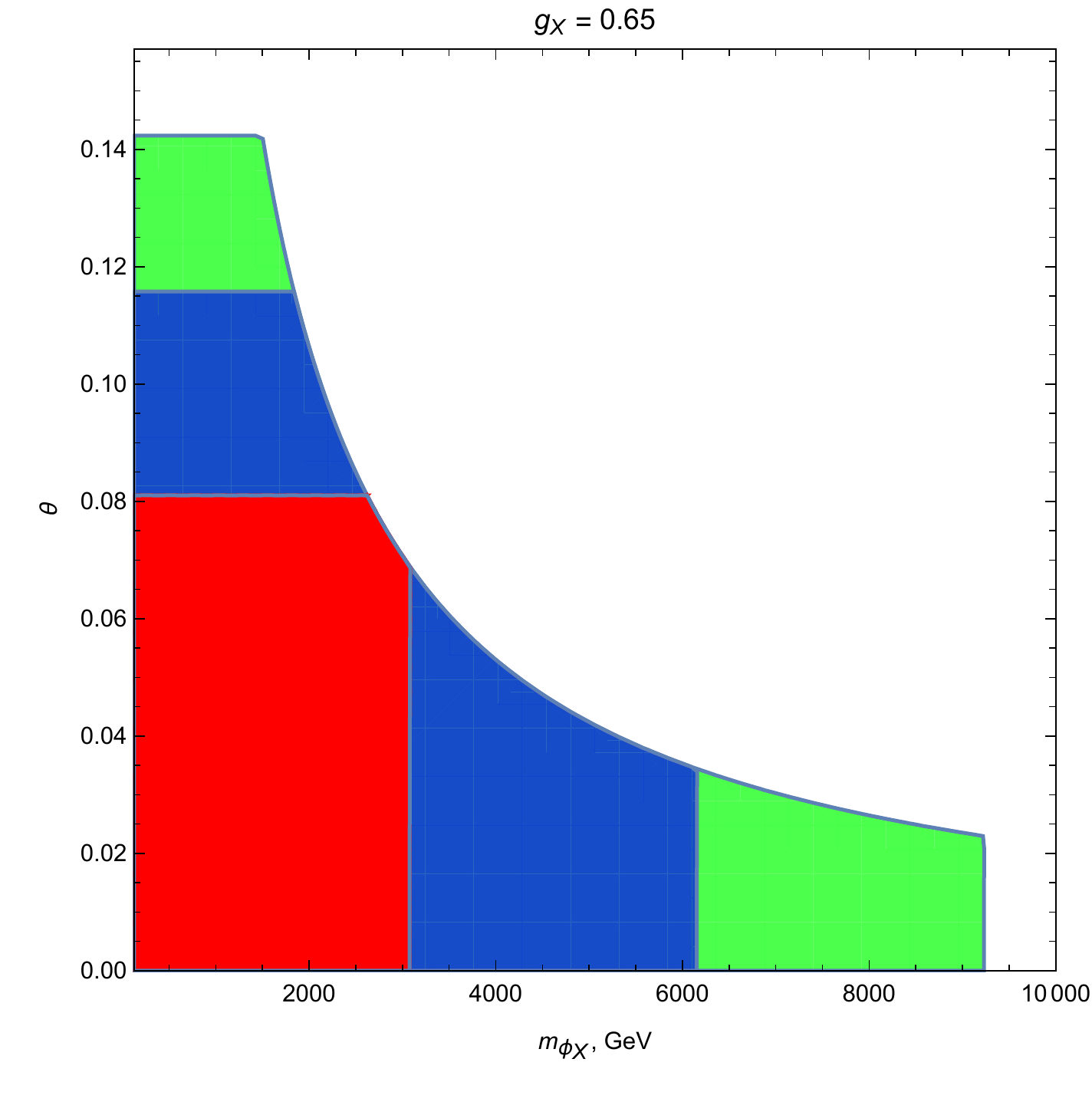}\hspace{60pt}
    \includegraphics[width=0.33\textwidth]{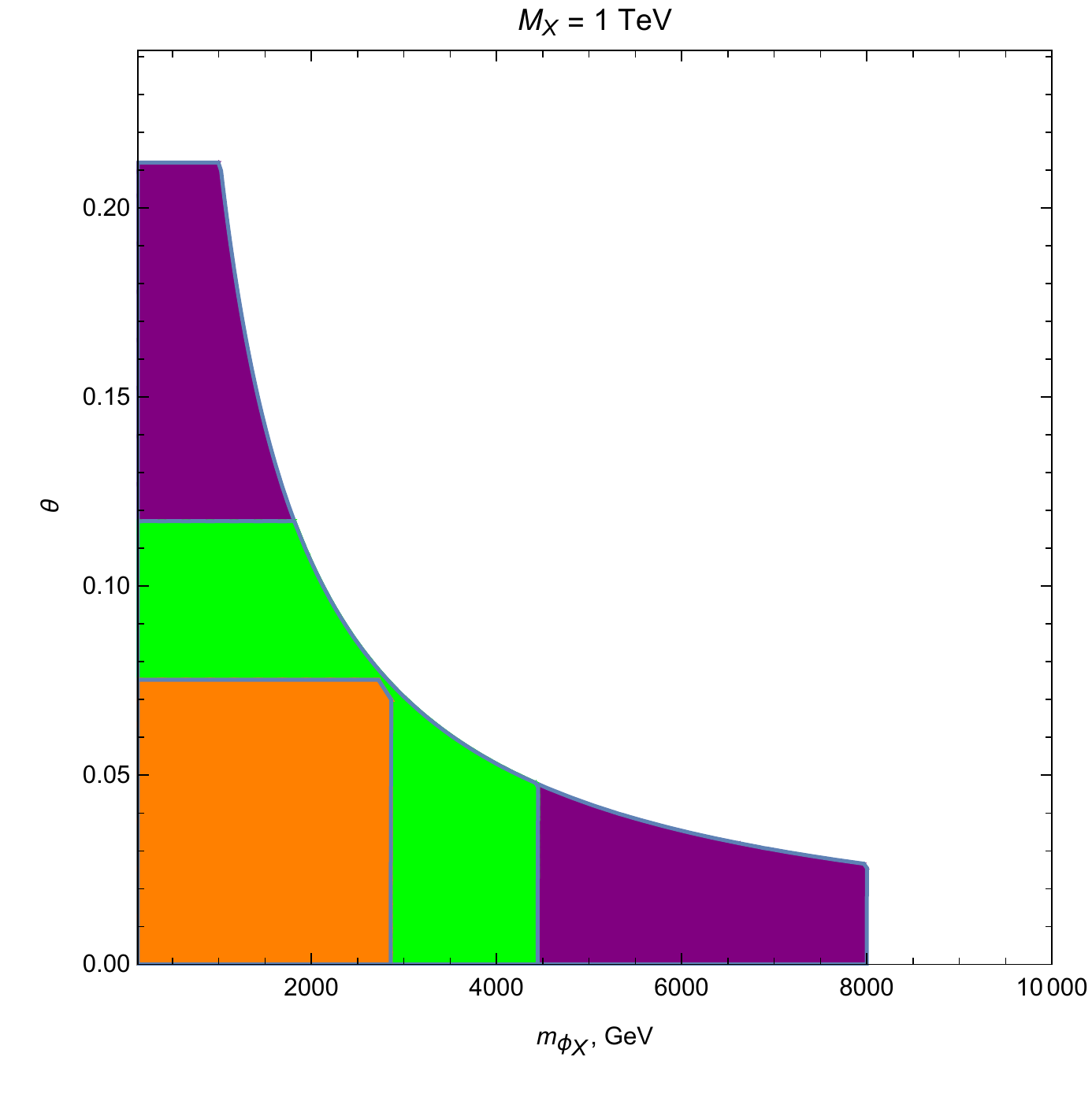}}
    \caption{The allowed parameter space of the heavy Higgs mass, $m_{\phi_X}$ vs the mixing angle, $\theta$ for (left figure) the DM mass, $M_X= 1$ TeV (red), $2$ TeV (blue) and $3$ TeV (green) with $SU(2)_{N}$ gauge coupling, $g_{X}=0.65$, and for (right figure) $g_{X}=0.25$ (purple), $0.45$ (green) and $0.7$ (orange) with $M_X=1$ TeV. Here we impose the constraints, $0\leq \lambda_{H,\phi,m}\leq 1$ and $\sigma_{\mathrm{SI}}\leq \sigma^{\mathrm{EXP}}_{\mathrm{SI}}$.}
    \label{mHvstheta}
\end{figure*}
\end{widetext}
From, Fig. \ref{mHvstheta}, we can see that the mixing angle is relatively small for smaller DM mass but there is no significant increase even when  we raise the DM mass while keeping the gauge coupling, $g_{X}$ fixed. Moreover, for smaller $g_{X}$, the allowed region is relatively larger when the DM mass is fixed. Nevertheless, for the region of parameter space where the muon and electron $g-2$ are relevant, $g_{X}$ is of the order $O(0.1-1)$ and $M_X$ is of the order $O(\mathrm{TeV})$, the allowed mixing angle between the SM Higgs and the BSM neutral Higgs remains quite small for a wide range of $m_{\phi_{X}}$.

\textbf{Dark Matter Relic Density}-- The relic abundance of $X$ is achieved via standard thermal freeze-out mechanism.  The $2\rightarrow 2$ (co)annihilation channels that give dominant contributions to the freeze-out of the non-relativistic $X$ are,

\begin{itemize}
    \item $X\,X^{\dagger}\rightarrow Z'Z'$ annihilation channel that  involves 4-point interaction, the exchange of $X$ in t and u channels and the exchange of $\phi_{X}$ and $h$ in the s-channel. As this annihilation mode consists of massive vector bosons in both initial and final states, the comparatively large multiplicities in this channel will lead to a larger cross-section. On the other hand, if the low-velocity approximation, $\sigma v =a+ b\, v^{2}$, is used to calculate the relic density, this annihilation channel turns out to be forbidden, which is not the case during the thermal freeze-out since it occurs at the temperature around $T_{f}\sim M_{X}/30 - M_{X}/20$, and the energy of the DM follows the Maxwell-Boltzmann distribution, as pointed out in \cite{Griest:1990kh}. As a result, this annihilation mode largely determines the relic abundance of the vector DM, $X$.
    
    \item Apart from $Z'\,Z'$ in the final states, one can also have $X\,X^{\dagger}\rightarrow Z'\,\phi_{X},\,\,\mathrm{and}\,\,Z'\,h$ i.e a vector boson and either the BSM neutral Higgs (when $M_X > m_{\phi_{X}}$)  or the SM higgs boson in the final state. This annihilation can proceed through the exchange of $Z'$ in the s channel and the exchange of $X$ at t and u channels.
    
    \item $X\,X^{\dagger}\rightarrow \overline{l}l,\,\overline{\nu}_{l}\nu_{l}$, i.e. to the SM charged lepton pairs and neutrino pairs ($l=e,\mu$ is the lepton flavor index) via the exchange of $E,\,E'$ and $N,\,N'$, respectively, in the $t$-channel and the exchange of $Z'$, $\phi_{X}$ and $h$ in the s channel.
    
    \item $X\,X^{\dagger}\rightarrow \phi_{X}\phi_{X},\,\phi_{X}\,h,\,\mathrm{and}\,\,h\,h$ i.e. annihilating into the pair of BSM neutral Higgs bosons and the SM Higgs bosons, and into one heavy Higgs and one SM Higgs bosons via the 4-point interaction, the exchange of $X$ in the t and u channels and the exchange of $\phi_X$ and $h$ in the s channel.
\end{itemize}

Besides, the additional channels that participate in the coannihilation are,
\begin{itemize}
    \item $X\,E,\,X\,E'\rightarrow \psi\psi'$, $X\,N,\,X\,N'\rightarrow \psi\psi'$ and their charge conjugated channels.
    \item $\overline{E}\,E,\,\overline{E'}\,E',\,\overline{E'}\,E\rightarrow \psi\,\psi'$ and $\overline{N}\,N,\,\overline{N'}\,N',\,\overline{N'}\,N\rightarrow \psi\,\psi'$ and their charge conjugated channels.
\end{itemize}
Because of large number of final states for these coannihilation channels, for simplicity we denote all of the allowed final states using $\psi\,\psi'$ where any one or both of $\psi,\,\psi'$ either indicate the SM particles or the particles carrying zero dark charge, i.e. $Z'$ and $\phi_{X}$ depending on kinematic conditions.

We calculate the relic abundance of $X$ using MicrOMEGAS v$\_\,5.2$ \cite{Belanger:2020gnr} in which we implement the model with the help of FeynRules \cite{Alloul:2013bka}. From Fig. \ref{mHvstheta}, we can see that the mixing angle between the SM and the BSM neutral Higgs, $\theta$ remains small for wide ranges of $m_{\phi_{X}}$, $M_X$ and $g_{X}$ values. Therefore, we have set $\theta\sim 10^{-4}$ for the subsequent calculation. For such small $\theta$, the annihilation channels that contain the interaction vertices with the SM Higgs and the dark sector particles, and vertices with the BSM neutral Higgs and the SM particles, will give negligible contribution to thermal freeze-out of the DM, $X$. Besides, for our region of interest, $M_X\geq m_{\phi_{X}}$, though the variation in the value of $m_{\phi_{X}}$ does not  significantly change the relic density of $X$, we vary $m_{\phi_{X}}$ within the range $126\,\mathrm{GeV}-0.9 M_{X}$ in our numerical study.

\section{Results}\label{RESULTsec}
In this section, we present our detailed numerical analysis and encapsulate  predictions of this theory. From aforementioned discussions, it is comprehensible that in this framework, the lepton AMMs and the dark matter physics are deeply intertwined with each other. 

\begin{figure}[th!]
\includegraphics[width=0.48\textwidth]{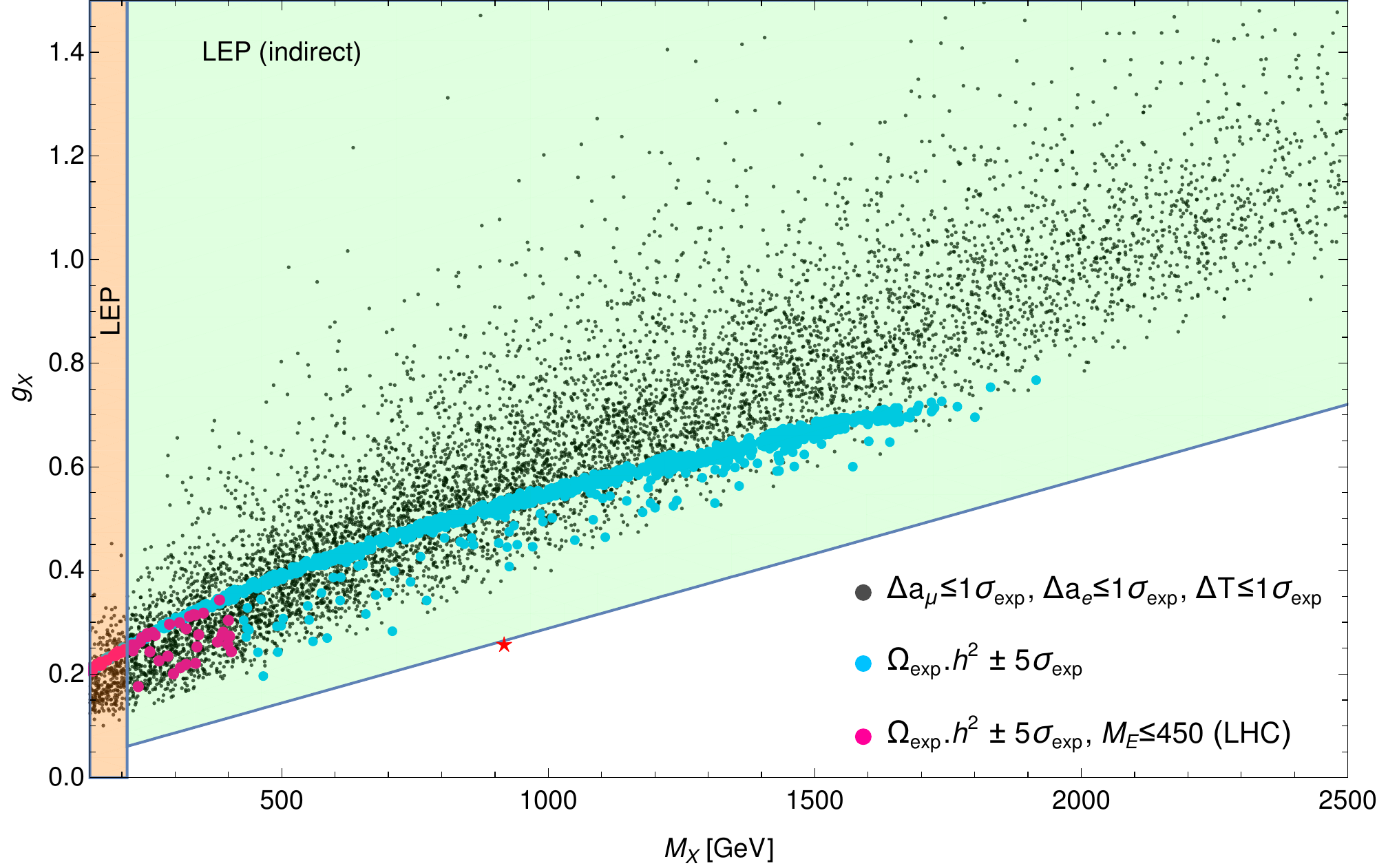}
\caption{The correlation between dark matter mass, $M_X$ and the $SU(2)_{X}$ gauge coupling, $g_{X}$. All points shown in the plot satisfy the electron and the muon anomalous magnetic moments within their $1\sigma$ experimental allowed values. Each point is also consistent with $T$-parameter constraints ($1\sigma$). Only the blue and red dots are consistent with dark matter relic abundance ($5\sigma$), however, points in red dots are ruled out by LHC search. Region shaded in orange is excluded by LEP direct search. Indirect search from LEP (shaded green region), however, rules out simultaneous explanation of both the electron and the muon AMMs within their expected $1\sigma$ values. For detailed explanation regarding the best-fit point, denoted by the red star, see text.   } \label{moneyplot}
\end{figure}

In our numerical analysis, we vary the relevant Yukawa couplings appearing in Eq. \eqref{yukawa} in the range $0.1 - 1$ for  diagonal entries and $0.01 - 1$ ($0.001 - 1$) for the off-diagonal entry in the muon (electron) sector. Since the main purpose of this work is to throw light on the electron and the muon AMMs, we do not include the associated tau sector in our numerical study. As for the gauge coupling and DM mass (we have treated $M_X$ to be the free parameter instead of $v_X$), the corresponding chosen ranges are $0.05 - 1.5$ and $140 - 5000$ GeV, respectively. By varying these parameters randomly within their above mentioned ranges for $10^8$ times, we compute the $T$-parameter, the muon and the electron AMMs, and dark matter relic abundance following the discussions of the previous sections. Our final result is depicted in Fig. \ref{moneyplot}.

In this Fig. \ref{moneyplot}, the plot shows the interdependence of the DM mass and the gauge coupling.  Points satisfying both $\Delta a_e$ and $\Delta a_\mu$ within their experimental $1\sigma$ values are shown in black, these points are also in agreement with $T$-parameter bounds within $1\sigma$.   However, the requirement of reproducing correct DM relic abundance \cite{Planck:2018vyg} rules out a large portion of the theory parameter space as can be seen from Fig. \ref{moneyplot}. Points that allow acceptable DM abundance are presented in blue  and red  colors. These red  points are further ruled out by the LHC  searches corresponding to the $M_E \leq 450$ GeV. Furthermore,  the parameter space ruled out by LEP direct and indirect searches  are shown in orange and green shaded regions, respectively. From this plot, it is evident that this model fails to explain both the electron and the muon $g-2$ within their $1\sigma$ experimental values,  due to several constraints arising mostly from LEP searches.

In search of finding a valid point in agreement with LEP as well as LHC limits, we perform a $\chi^2$ analysis. In this numerical procedure, we minimize the function $\chi^2=\sum_i P^2_i$, where the pull is defined as $P_i=(T_i-E_i)/\sigma_i$. For an observable $i$, $T_i, E_i$, and $\sigma_i$ denote theory prediction, experimental central value, and experimental $1\sigma$ uncertainty, respectively.  The sum is taken over $i=\{\Delta a_e, \Delta a_\mu, \Delta T, \Omega. h^2\}$. This is a constrained optimization that includes the experimental constraints implemented on top of  $\chi^2$-function. The best-fit point obtained in this procedure leads to $\Delta a_\mu= 1.5  \times 10^{-9}$, which corresponds to a pull of $P=-1.69$ ($\chi^2_{total}=3.8$). For rest of the observables pulls are smaller than unity.  This best-fit point is allowed by all bounds arising from current experiments, as can be seen from Fig. \ref{moneyplot} (point marked as red star; for this best-fit $M_X/(g_X/2)= 7.19$ TeV $>\Lambda_{LEP}$). Despite of satisfying LEP bound, since the muon AMM cannot be fitted within its $1\sigma$ range, we conclude that this model disfavors a simultaneous explanation of both the muon and the electron $g-2$. A behavior of this type can be understood from Fig. \ref{gM2}, which demonstrates that for a fixed DM mass, $(g-2)_\mu$ demands higher value of $g_X$ (compared to correctly reproducing $(g-2)_e$ within $1\sigma$ that allows smaller values of $g_X$), hence conflicting with LEP bound.

\begin{figure}[th!]
\includegraphics[width=0.48\textwidth]{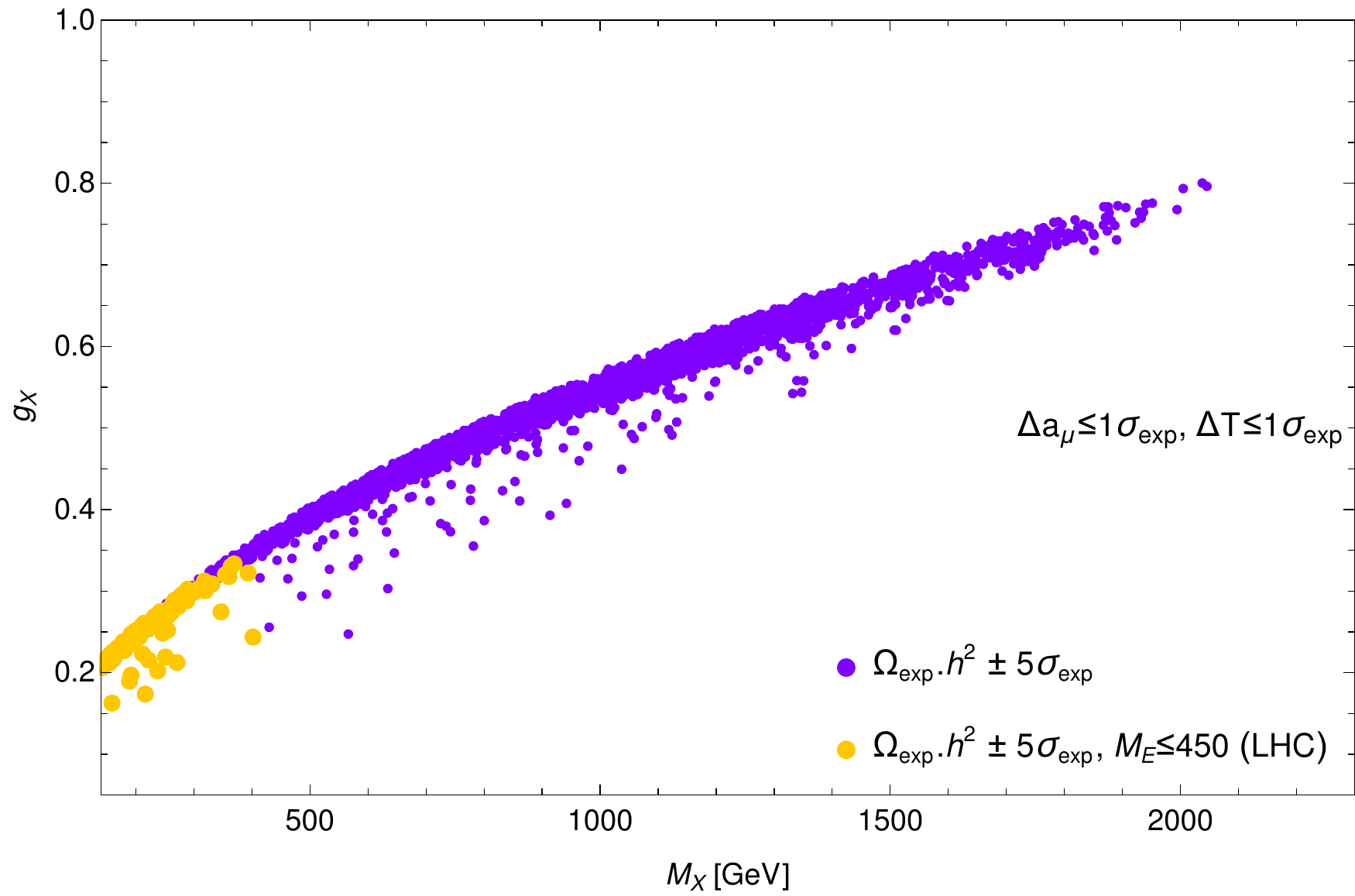}
\caption{The correlation between dark matter mass, $M_X$ and the $SU(2)_{X}$ gauge coupling, $g_{X}$. All points shown in this plot satisfy the muon anomalous magnetic moment and dark matter relic abundance within their $1\sigma$ and $5\sigma$  ranges, respectively. Each point is also consistent with $T$-parameter constraints ($1\sigma$). Yellow points are ruled out by LHC searches. For detailed, see text.} \label{moneyplotMUON}
\end{figure}

The only way to overcome the stringent LEP bound is to forbid the electron to couple to $Z^{\prime}$; in the following we explore such a possibility. This scenario is equivalent to  having a set of fermions listed in Eq. \eqref{F1}-\eqref{F2} only for the muon sector. The first and the third generation of leptons are then identical to that of the SM. LEP bounds are no longer present,  and the full parameter space consistent with the muon $g-2$ and DM relic abundance is presented in Fig. \ref{moneyplotMUON}. To generate this plot, we follow the same procedure as that of Fig \ref{moneyplot}.  As can be seen from Fig.\ \ref{moneyplotMUON}, when the muon AMM and the DM relic density constraints are combined with the assumption that $m_{\phi_X} < M_X$, the allowed parameter space of this model is rather limited.  This corresponds to gauge coupling $g_X\sim 0.2 - 0.8$ and DM mass $M_X\sim 500 - 2000$ GeV. The lower limit of the DM mass $M_X \gtrsim 0.5$ TeV is fixed by LHC searches, whereas the upper limit $M_X \lesssim 2$ TeV is restricted by the viability of reproducing correct $\Delta a_\mu$. The reason that both Fig.  \ref{moneyplot} (that includes both $(g-2)_\mu$ and $(g-2)_e$) and Fig. \ref{moneyplotMUON} that includes only $(g-2)_\mu$) have similar cut-off from the lower side can be clearly understood from Fig. \ref{gM2}.  In addition, this muon-specific scenario can be explored in the upcoming Muon collider \cite{Muoncollider}.

Furthermore, when included, the tau sector will contribute to the thermal freeze-out of the DM, however, such effects would be negligible because the freeze-out process is completely dominated by the DM annihilating into $Z^\prime Z^\prime$ channel for our preferred region of parameter space.

\section{Conclusion}\label{CONsec}
In this work, we have presented a model that sheds light on the origin of the dark matter and also resolves tantalizing anomaly observed in the muon anomalous magnetic moment.  The proposed framework extends the SM by $SU(2)_X$ gauge symmetry, under which SM leptons transform non-trivially.    The new gauge bosons that play the role of dark matter, with the help of additional fermions needed for anomaly cancellation, furnish prescribed quantum corrections towards the lepton anomalous magnetic moments. When  contemporary collider constraints and EW precision measurements are taken into consideration,  we find that 
simultaneous explanation of the muon and the electron AMMs along with obtaining right DM relic abundance is highly disfavored.  This leads us to a specific scenario for which only the muon is charged under the added $SU(2)_X$. In such a scenario, our analysis shows that  a  viable parameter space of the model  for which the new gauge coupling is $g_X\sim 0.2 - 0.8$ and DM mass is $M_X\sim 0.5 - 2$ TeV can explain dark matter relic abundance as well as the large deviation observed in the muon $g-2$, recently confirmed by the FNAL result.

\begin{acknowledgments}
{\textbf {\textit {Acknowledgments.--}}} The work of S.S.\ has been supported by the Swiss National Science Foundation.
\end{acknowledgments}

\bibliographystyle{style}
\bibliography{reference}

\providecommand{\href}[2]{#2}\begingroup\raggedright\begin{thebibliography}{10}

\bibitem{Oort}
J.~Oort, ``{The Force Exerted by the Stellar System in the Direction
  Perpendicular to the Galactic Plane and Some Related Problems},'' {\em
  Bulletin of the Astronomical Institutes of the Netherlands} {\bfseries 6}
  (1932) 249--287.

\bibitem{Zwicky:1933gu}
F.~Zwicky, ``{Die Rotverschiebung von extragalaktischen Nebeln},''
  \href{http://dx.doi.org/10.1007/s10714-008-0707-4}{{\em Helv. Phys. Acta}
  {\bfseries 6} (1933) 110--127}.

\bibitem{Farrar:2017eqq}
G.~R. Farrar, ``{Stable Sexaquark},''
  \href{http://arxiv.org/abs/1708.08951}{{\ttfamily arXiv:1708.08951
  [hep-ph]}}.

\bibitem{Gross:2018ivp}
C.~Gross, A.~Polosa, A.~Strumia, A.~Urbano, and W.~Xue, ``{Dark Matter in the
  Standard Model?},'' \href{http://dx.doi.org/10.1103/PhysRevD.98.063005}{{\em
  Phys. Rev. D} {\bfseries 98} no.~6, (2018) 063005},
  \href{http://arxiv.org/abs/1803.10242}{{\ttfamily arXiv:1803.10242
  [hep-ph]}}.

\bibitem{London:1986dk}
D.~London and J.~L. Rosner, ``{Extra Gauge Bosons in E(6)},''
  \href{http://dx.doi.org/10.1103/PhysRevD.34.1530}{{\em Phys. Rev. D}
  {\bfseries 34} (1986) 1530}.

\bibitem{DiazCruz:2010dc}
J.~L. Diaz-Cruz and E.~Ma, ``{Neutral SU(2) Gauge Extension of the Standard
  Model and a Vector-Boson Dark-Matter Candidate},''
  \href{http://dx.doi.org/10.1016/j.physletb.2010.11.039}{{\em Phys. Lett. B}
  {\bfseries 695} (2011) 264--267},
  \href{http://arxiv.org/abs/1007.2631}{{\ttfamily arXiv:1007.2631 [hep-ph]}}.

\bibitem{Bhattacharya:2011tr}
S.~Bhattacharya, J.~L. Diaz-Cruz, E.~Ma, and D.~Wegman, ``{Dark
  Vector-Gauge-Boson Model},''
  \href{http://dx.doi.org/10.1103/PhysRevD.85.055008}{{\em Phys. Rev. D}
  {\bfseries 85} (2012) 055008},
  \href{http://arxiv.org/abs/1107.2093}{{\ttfamily arXiv:1107.2093 [hep-ph]}}.

\bibitem{Ma:2012xj}
E.~Ma and J.~Wudka, ``{Vector-Boson-Induced Neutrino Mass},''
  \href{http://dx.doi.org/10.1016/j.physletb.2012.05.008}{{\em Phys. Lett. B}
  {\bfseries 712} (2012) 391--395},
  \href{http://arxiv.org/abs/1202.3098}{{\ttfamily arXiv:1202.3098 [hep-ph]}}.

\bibitem{Davoudiasl:2013jma}
H.~Davoudiasl and I.~M. Lewis, ``{Dark Matter from Hidden Forces},''
  \href{http://dx.doi.org/10.1103/PhysRevD.89.055026}{{\em Phys. Rev. D}
  {\bfseries 89} no.~5, (2014) 055026},
  \href{http://arxiv.org/abs/1309.6640}{{\ttfamily arXiv:1309.6640 [hep-ph]}}.

\bibitem{Fornal:2017owa}
B.~Fornal, Y.~Shirman, T.~M.~P. Tait, and J.~R. West, ``{Asymmetric dark matter
  and baryogenesis from $SU(2)_{\ell}$},''
  \href{http://dx.doi.org/10.1103/PhysRevD.96.035001}{{\em Phys. Rev. D}
  {\bfseries 96} no.~3, (2017) 035001},
  \href{http://arxiv.org/abs/1703.00199}{{\ttfamily arXiv:1703.00199
  [hep-ph]}}.

\bibitem{Ma:2021roh}
E.~Ma, ``{Non-Abelian gauge lepton symmetry as the gateway to dark matter},''
  \href{http://dx.doi.org/10.1016/j.physletb.2021.136456}{{\em Phys. Lett. B}
  {\bfseries 819} (2021) 136456},
  \href{http://arxiv.org/abs/2105.04466}{{\ttfamily arXiv:2105.04466
  [hep-ph]}}.

\bibitem{Bennett:2006fi}
{\bfseries Muon g-2} Collaboration, G.~W. Bennett {\em et~al.}, ``{Final Report
  of the Muon E821 Anomalous Magnetic Moment Measurement at BNL},''
  \href{http://dx.doi.org/10.1103/PhysRevD.73.072003}{{\em Phys. Rev.}
  {\bfseries D73} (2006) 072003},
\href{http://arxiv.org/abs/hep-ex/0602035}{{\ttfamily arXiv:hep-ex/0602035
  [hep-ex]}}.

\bibitem{Abi:2021gix}
{\bfseries Muon g-2} Collaboration, B.~Abi {\em et~al.}, ``{Measurement of the
  Positive Muon Anomalous Magnetic Moment to 0.46~ppm},''
  \href{http://dx.doi.org/10.1103/PhysRevLett.126.141801}{{\em Phys. Rev.
  Lett.} {\bfseries 126} no.~14, (2021) 141801},
  \href{http://arxiv.org/abs/2104.03281}{{\ttfamily arXiv:2104.03281
  [hep-ex]}}.

\bibitem{Aoyama:2020ynm}
T.~Aoyama {\em et~al.}, ``{The anomalous magnetic moment of the muon in the
  Standard Model},''
  \href{http://dx.doi.org/10.1016/j.physrep.2020.07.006}{{\em Phys. Rept.}
  {\bfseries 887} (2020) 1--166},
  \href{http://arxiv.org/abs/2006.04822}{{\ttfamily arXiv:2006.04822
  [hep-ph]}}.

\bibitem{Athron:2021iuf}
P.~Athron, C.~Bal\'azs, D.~H. Jacob, W.~Kotlarski, D.~St\"ockinger, and
  H.~St\"ockinger-Kim, ``{New physics explanations of $a_\mu$ in light of the
  FNAL muon $g-2$ measurement},''
  \href{http://arxiv.org/abs/2104.03691}{{\ttfamily arXiv:2104.03691
  [hep-ph]}}.

\bibitem{Parker:2018vye}
R.~H. Parker, C.~Yu, W.~Zhong, B.~Estey, and H.~Mueller, ``{Measurement of the
  fine-structure constant as a test of the Standard Model},''
  \href{http://dx.doi.org/10.1126/science.aap7706}{{\em Science} {\bfseries
  360} (2018) 191},
\href{http://arxiv.org/abs/1812.04130}{{\ttfamily arXiv:1812.04130
  [physics.atom-ph]}}.

\bibitem{Morel:2020dww}
L.~Morel, Z.~Yao, P.~Cladé, and S.~Guellati-Khélifa, ``{Determination of the
  fine-structure constant with an accuracy of 81 parts per trillion},''
\href{http://dx.doi.org/10.1038/s41586-020-2964-7}{{\em Nature} {\bfseries 588}
  no.~7836, (2020) 61--65}.

\bibitem{Aoyama:2017uqe}
T.~Aoyama, T.~Kinoshita, and M.~Nio, ``{Revised and Improved Value of the QED
  Tenth-Order Electron Anomalous Magnetic Moment},''
  \href{http://dx.doi.org/10.1103/PhysRevD.97.036001}{{\em Phys. Rev.}
  {\bfseries D97} no.~3, (2018) 036001},
\href{http://arxiv.org/abs/1712.06060}{{\ttfamily arXiv:1712.06060 [hep-ph]}}.

\bibitem{Giudice:2012ms}
G.~F. Giudice, P.~Paradisi, and M.~Passera, ``{Testing new physics with the
  electron g-2},'' \href{http://dx.doi.org/10.1007/JHEP11(2012)113}{{\em JHEP}
  {\bfseries 11} (2012) 113},
\href{http://arxiv.org/abs/1208.6583}{{\ttfamily arXiv:1208.6583 [hep-ph]}}.

\bibitem{Davoudiasl:2018fbb}
H.~Davoudiasl and W.~J. Marciano, ``{Tale of two anomalies},''
  \href{http://dx.doi.org/10.1103/PhysRevD.98.075011}{{\em Phys. Rev.}
  {\bfseries D98} no.~7, (2018) 075011},
\href{http://arxiv.org/abs/1806.10252}{{\ttfamily arXiv:1806.10252 [hep-ph]}}.

\bibitem{Crivellin:2018qmi}
A.~Crivellin, M.~Hoferichter, and P.~Schmidt-Wellenburg, ``{Combined
  explanations of $(g-2)_{\mu,e}$ and implications for a large muon EDM},''
  \href{http://dx.doi.org/10.1103/PhysRevD.98.113002}{{\em Phys. Rev.}
  {\bfseries D98} no.~11, (2018) 113002},
\href{http://arxiv.org/abs/1807.11484}{{\ttfamily arXiv:1807.11484 [hep-ph]}}.

\bibitem{Liu:2018xkx}
J.~Liu, C.~E.~M. Wagner, and X.-P. Wang, ``{A light complex scalar for the
  electron and muon anomalous magnetic moments},''
  \href{http://dx.doi.org/10.1007/JHEP03(2019)008}{{\em JHEP} {\bfseries 03}
  (2019) 008},
\href{http://arxiv.org/abs/1810.11028}{{\ttfamily arXiv:1810.11028 [hep-ph]}}.

\bibitem{Dutta:2018fge}
B.~Dutta and Y.~Mimura, ``{Electron $g-2$ with flavor violation in MSSM},''
  \href{http://dx.doi.org/10.1016/j.physletb.2018.12.070}{{\em Phys. Lett.}
  {\bfseries B790} (2019) 563--567},
\href{http://arxiv.org/abs/1811.10209}{{\ttfamily arXiv:1811.10209 [hep-ph]}}.

\bibitem{Han:2018znu}
X.-F. Han, T.~Li, L.~Wang, and Y.~Zhang, ``{Simple interpretations of lepton
  anomalies in the lepton-specific inert two-Higgs-doublet model},''
  \href{http://dx.doi.org/10.1103/PhysRevD.99.095034}{{\em Phys. Rev.}
  {\bfseries D99} no.~9, (2019) 095034},
\href{http://arxiv.org/abs/1812.02449}{{\ttfamily arXiv:1812.02449 [hep-ph]}}.

\bibitem{Crivellin:2019mvj}
A.~Crivellin and M.~Hoferichter, ``{Combined explanations of $(g-2)_\mu$,
  $(g-2)_e$ and implications for a large muon EDM},'' in {\em {33rd Rencontres
  de Physique de La Vallée d'Aoste (LaThuile 2019) La Thuile, Aosta, Italy,
  March 10-16, 2019}}.
\newblock 2019.
\newblock
\href{http://arxiv.org/abs/1905.03789}{{\ttfamily arXiv:1905.03789 [hep-ph]}}.
\newblock

\bibitem{Endo:2019bcj}
M.~Endo and W.~Yin, ``{Explaining electron and muon $g-2$ anomaly in SUSY
  without lepton-flavor mixings},''
  \href{http://dx.doi.org/10.1007/JHEP08(2019)122}{{\em JHEP} {\bfseries 08}
  (2019) 122},
\href{http://arxiv.org/abs/1906.08768}{{\ttfamily arXiv:1906.08768 [hep-ph]}}.

\bibitem{Abdullah:2019ofw}
M.~Abdullah, B.~Dutta, S.~Ghosh, and T.~Li, ``{$(g-2)_{\mu,e}$ and the ANITA
  anomalous events in a three-loop neutrino mass model},''
  \href{http://dx.doi.org/10.1103/PhysRevD.100.115006}{{\em Phys. Rev.}
  {\bfseries D100} no.~11, (2019) 115006},
\href{http://arxiv.org/abs/1907.08109}{{\ttfamily arXiv:1907.08109 [hep-ph]}}.

\bibitem{Bauer:2019gfk}
M.~Bauer, M.~Neubert, S.~Renner, M.~Schnubel, and A.~Thamm, ``{Axion-like
  particles, lepton-flavor violation and a new explanation of $a_\mu$ and
  $a_e$},''
\href{http://arxiv.org/abs/1908.00008}{{\ttfamily arXiv:1908.00008 [hep-ph]}}.

\bibitem{Badziak:2019gaf}
M.~Badziak and K.~Sakurai, ``{Explanation of electron and muon g − 2
  anomalies in the MSSM},''
  \href{http://dx.doi.org/10.1007/JHEP10(2019)024}{{\em JHEP} {\bfseries 10}
  (2019) 024},
\href{http://arxiv.org/abs/1908.03607}{{\ttfamily arXiv:1908.03607 [hep-ph]}}.

\bibitem{Hiller:2019mou}
G.~Hiller, C.~Hormigos-Feliu, D.~F. Litim, and T.~Steudtner, ``{Anomalous
  magnetic moments from asymptotic safety},''
\href{http://arxiv.org/abs/1910.14062}{{\ttfamily arXiv:1910.14062 [hep-ph]}}.

\bibitem{CarcamoHernandez:2019ydc}
A.~E. Cárcamo~Hernández, S.~F. King, H.~Lee, and S.~J. Rowley, ``{Is it
  possible to explain the muon and electron $g-2$ in a $Z^{\prime}$ model?},''
\href{http://arxiv.org/abs/1910.10734}{{\ttfamily arXiv:1910.10734 [hep-ph]}}.

\bibitem{Cornella:2019uxs}
C.~Cornella, P.~Paradisi, and O.~Sumensari, ``{Hunting for ALPs with Lepton
  Flavor Violation},''
\href{http://arxiv.org/abs/1911.06279}{{\ttfamily arXiv:1911.06279 [hep-ph]}}.

\bibitem{Endo:2020mev}
M.~Endo, S.~Iguro, and T.~Kitahara, ``{Probing $e\mu$ flavor-violating ALP at
  Belle II},''
\href{http://arxiv.org/abs/2002.05948}{{\ttfamily arXiv:2002.05948 [hep-ph]}}.

\bibitem{CarcamoHernandez:2020pxw}
A.~E. Cárcamo~Hernández, Y.~H. Velásquez, S.~Kovalenko, H.~N. Long, N.~A.
  Pérez-Julve, and V.~V. Vien, ``{Fermion masses and mixings and $g-2$
  anomalies in a low scale 3-3-1 model},''
\href{http://arxiv.org/abs/2002.07347}{{\ttfamily arXiv:2002.07347 [hep-ph]}}.

\bibitem{Haba:2020gkr}
N.~Haba, Y.~Shimizu, and T.~Yamada, ``{Muon and Electron $g-2$ and the Origin
  of Fermion Mass Hierarchy},''
\href{http://arxiv.org/abs/2002.10230}{{\ttfamily arXiv:2002.10230 [hep-ph]}}.

\bibitem{Bigaran:2020jil}
I.~Bigaran and R.~R. Volkas, ``{Getting chirality right: top-philic scalar
  leptoquark solution to the $(g-2)_{e,\mu}$ puzzle},''
\href{http://arxiv.org/abs/2002.12544}{{\ttfamily arXiv:2002.12544 [hep-ph]}}.

\bibitem{Jana:2020pxx}
S.~Jana, V.~P. K., and S.~Saad, ``{Resolving electron and muon $g-2$ within the
  2HDM},'' \href{http://dx.doi.org/10.1103/PhysRevD.101.115037}{{\em Phys. Rev.
  D} {\bfseries 101} no.~11, (2020) 115037},
  \href{http://arxiv.org/abs/2003.03386}{{\ttfamily arXiv:2003.03386
  [hep-ph]}}.

\bibitem{Calibbi:2020emz}
L.~Calibbi, M.~L. L\'opez-Ib\'a\~nez, A.~Melis, and O.~Vives, ``{Muon and
  electron $g−2$ and lepton masses in flavor models},''
  \href{http://dx.doi.org/10.1007/JHEP06(2020)087}{{\em JHEP} {\bfseries 06}
  (2020) 087}, \href{http://arxiv.org/abs/2003.06633}{{\ttfamily
  arXiv:2003.06633 [hep-ph]}}.

\bibitem{Chen:2020jvl}
C.-H. Chen and T.~Nomura, ``{Electron and muon $g-2$, radiative neutrino mass,
  and $\ell' \to \ell \gamma$ in a $U(1)_{e-\mu}$ model},''
  \href{http://dx.doi.org/10.1016/j.nuclphysb.2021.115314}{{\em Nucl. Phys. B}
  {\bfseries 964} (2021) 115314},
  \href{http://arxiv.org/abs/2003.07638}{{\ttfamily arXiv:2003.07638
  [hep-ph]}}.

\bibitem{Yang:2020bmh}
J.-L. Yang, T.-F. Feng, and H.-B. Zhang, ``{Electron and muon $(g-2)$ in the
  B-LSSM},'' \href{http://dx.doi.org/10.1088/1361-6471/ab7986}{{\em J. Phys. G}
  {\bfseries 47} no.~5, (2020) 055004},
  \href{http://arxiv.org/abs/2003.09781}{{\ttfamily arXiv:2003.09781
  [hep-ph]}}.

\bibitem{Hati:2020fzp}
C.~Hati, J.~Kriewald, J.~Orloff, and A.~M. Teixeira, ``{Anomalies in $^8$Be
  nuclear transitions and $(g-2)_{e,\mu}$: towards a minimal combined
  explanation},'' \href{http://dx.doi.org/10.1007/JHEP07(2020)235}{{\em JHEP}
  {\bfseries 07} (2020) 235}, \href{http://arxiv.org/abs/2005.00028}{{\ttfamily
  arXiv:2005.00028 [hep-ph]}}.

\bibitem{Dutta:2020scq}
B.~Dutta, S.~Ghosh, and T.~Li, ``{Explaining $(g-2)_{\mu,e}$, the KOTO anomaly
  and the MiniBooNE excess in an extended Higgs model with sterile
  neutrinos},'' \href{http://dx.doi.org/10.1103/PhysRevD.102.055017}{{\em Phys.
  Rev. D} {\bfseries 102} no.~5, (2020) 055017},
  \href{http://arxiv.org/abs/2006.01319}{{\ttfamily arXiv:2006.01319
  [hep-ph]}}.

\bibitem{Botella:2020xzf}
F.~J. Botella, F.~Cornet-Gomez, and M.~Nebot, ``{Electron and muon $g-2$
  anomalies in general flavour conserving two Higgs doublets models},''
  \href{http://dx.doi.org/10.1103/PhysRevD.102.035023}{{\em Phys. Rev. D}
  {\bfseries 102} no.~3, (2020) 035023},
  \href{http://arxiv.org/abs/2006.01934}{{\ttfamily arXiv:2006.01934
  [hep-ph]}}.

\bibitem{Chen:2020tfr}
K.-F. Chen, C.-W. Chiang, and K.~Yagyu, ``{An explanation for the muon and
  electron $g − 2$ anomalies and dark matter},''
  \href{http://dx.doi.org/10.1007/JHEP09(2020)119}{{\em JHEP} {\bfseries 09}
  (2020) 119}, \href{http://arxiv.org/abs/2006.07929}{{\ttfamily
  arXiv:2006.07929 [hep-ph]}}.

\bibitem{Dorsner:2020aaz}
I.~Dor\v{s}ner, S.~Fajfer, and S.~Saad, ``{$\mu \to e \gamma$ selecting scalar
  leptoquark solutions for the $(g-2)_{e,\mu}$ puzzles},''
  \href{http://dx.doi.org/10.1103/PhysRevD.102.075007}{{\em Phys. Rev. D}
  {\bfseries 102} no.~7, (2020) 075007},
  \href{http://arxiv.org/abs/2006.11624}{{\ttfamily arXiv:2006.11624
  [hep-ph]}}.

\bibitem{Arbelaez:2020rbq}
C.~Arbel\'aez, R.~Cepedello, R.~M. Fonseca, and M.~Hirsch, ``{$(g-2)$ anomalies
  and neutrino mass},''
  \href{http://dx.doi.org/10.1103/PhysRevD.102.075005}{{\em Phys. Rev. D}
  {\bfseries 102} no.~7, (2020) 075005},
  \href{http://arxiv.org/abs/2007.11007}{{\ttfamily arXiv:2007.11007
  [hep-ph]}}.

\bibitem{Jana:2020joi}
S.~Jana, P.~K. Vishnu, W.~Rodejohann, and S.~Saad, ``{Dark matter assisted
  lepton anomalous magnetic moments and neutrino masses},''
  \href{http://dx.doi.org/10.1103/PhysRevD.102.075003}{{\em Phys. Rev.}
  {\bfseries D102} no.~7, (2020) 075003},
\href{http://arxiv.org/abs/2008.02377}{{\ttfamily arXiv:2008.02377 [hep-ph]}}.

\bibitem{Chua:2020dya}
C.-K. Chua, ``{Data-driven study of the implications of anomalous magnetic
  moments and lepton flavor violating processes of $e$, $\mu$ and $\tau$},''
  \href{http://dx.doi.org/10.1103/PhysRevD.102.055022}{{\em Phys. Rev. D}
  {\bfseries 102} no.~5, (2020) 055022},
  \href{http://arxiv.org/abs/2004.11031}{{\ttfamily arXiv:2004.11031
  [hep-ph]}}.

\bibitem{Chun:2020uzw}
E.~J. Chun and T.~Mondal, ``{Explaining $g-2$ anomalies in two Higgs doublet
  model with vector-like leptons},''
  \href{http://dx.doi.org/10.1007/JHEP11(2020)077}{{\em JHEP} {\bfseries 11}
  (2020) 077}, \href{http://arxiv.org/abs/2009.08314}{{\ttfamily
  arXiv:2009.08314 [hep-ph]}}.

\bibitem{Li:2020dbg}
S.-P. Li, X.-Q. Li, Y.-Y. Li, Y.-D. Yang, and X.~Zhang, ``{Power-aligned 2HDM:
  a correlative perspective on $(g-2)_{e,\mu}$},''
  \href{http://dx.doi.org/10.1007/JHEP01(2021)034}{{\em JHEP} {\bfseries 01}
  (2021) 034}, \href{http://arxiv.org/abs/2010.02799}{{\ttfamily
  arXiv:2010.02799 [hep-ph]}}.

\bibitem{DelleRose:2020oaa}
L.~Delle~Rose, S.~Khalil, and S.~Moretti, ``{Explaining electron and muon $g$
  \ensuremath{-} 2 anomalies in an Aligned 2-Higgs Doublet Model with
  right-handed neutrinos},''
  \href{http://dx.doi.org/10.1016/j.physletb.2021.136216}{{\em Phys. Lett. B}
  {\bfseries 816} (2021) 136216},
  \href{http://arxiv.org/abs/2012.06911}{{\ttfamily arXiv:2012.06911
  [hep-ph]}}.

\bibitem{Kowalska:2020zve}
K.~Kowalska and E.~M. Sessolo, ``{Minimal models for g-2 and dark matter
  confront asymptotic safety},''
  \href{http://dx.doi.org/10.1103/PhysRevD.103.115032}{{\em Phys. Rev. D}
  {\bfseries 103} no.~11, (2021) 115032},
  \href{http://arxiv.org/abs/2012.15200}{{\ttfamily arXiv:2012.15200
  [hep-ph]}}.

\bibitem{Hernandez:2021tii}
A.~E.~C. Hern\'andez, S.~F. King, and H.~Lee, ``{Fermion mass hierarchies from
  vectorlike families with an extended 2HDM and a possible explanation for the
  electron and muon anomalous magnetic moments},''
  \href{http://dx.doi.org/10.1103/PhysRevD.103.115024}{{\em Phys. Rev. D}
  {\bfseries 103} no.~11, (2021) 115024},
  \href{http://arxiv.org/abs/2101.05819}{{\ttfamily arXiv:2101.05819
  [hep-ph]}}.

\bibitem{Bodas:2021fsy}
A.~Bodas, R.~Coy, and S.~J.~D. King, ``{Solving the electron and muon $g-2$
  anomalies in $Z'$ models},''
  \href{http://arxiv.org/abs/2102.07781}{{\ttfamily arXiv:2102.07781
  [hep-ph]}}.

\bibitem{Cao:2021lmj}
J.~Cao, Y.~He, J.~Lian, D.~Zhang, and P.~Zhu, ``{Electron and Muon Anomalous
  Magnetic Moments in the Inverse Seesaw Extended NMSSM},''
  \href{http://arxiv.org/abs/2102.11355}{{\ttfamily arXiv:2102.11355
  [hep-ph]}}.

\bibitem{Mondal:2021vou}
T.~Mondal and H.~Okada, ``{Inverse seesaw and $(g-2)$ anomalies in $B-L$
  extended two Higgs doublet model},''
  \href{http://arxiv.org/abs/2103.13149}{{\ttfamily arXiv:2103.13149
  [hep-ph]}}.

\bibitem{CarcamoHernandez:2021iat}
A.~E. C\'arcamo~Hern\'andez, C.~Espinoza, J.~Carlos G\'omez-Izquierdo, and
  M.~Mondrag\'on, ``{Fermion masses and mixings, dark matter, leptogenesis and
  $g-2$ muon anomaly in an extended 2HDM with inverse seesaw},''
  \href{http://arxiv.org/abs/2104.02730}{{\ttfamily arXiv:2104.02730
  [hep-ph]}}.

\bibitem{Han:2021gfu}
X.-F. Han, T.~Li, H.-X. Wang, L.~Wang, and Y.~Zhang, ``{Lepton-specific inert
  two-Higgs-doublet model confronted with the new results for muon and electron
  g-2 anomalies and multi-lepton searches at the LHC},''
  \href{http://arxiv.org/abs/2104.03227}{{\ttfamily arXiv:2104.03227
  [hep-ph]}}.

\bibitem{Escribano:2021css}
P.~Escribano, J.~Terol-Calvo, and A.~Vicente, ``{$\boldsymbol{(g-2)_{e,\mu}}$
  in an extended inverse type-III seesaw model},''
  \href{http://dx.doi.org/10.1103/PhysRevD.103.115018}{{\em Phys. Rev. D}
  {\bfseries 103} no.~11, (2021) 115018},
  \href{http://arxiv.org/abs/2104.03705}{{\ttfamily arXiv:2104.03705
  [hep-ph]}}.

\bibitem{CarcamoHernandez:2021qhf}
A.~E. C\'arcamo~Hern\'andez, S.~Kovalenko, M.~Maniatis, and I.~Schmidt,
  ``{Fermion mass hierarchy and g-2 anomalies in an extended 3HDM Model},''
  \href{http://arxiv.org/abs/2104.07047}{{\ttfamily arXiv:2104.07047
  [hep-ph]}}.

\bibitem{Chang:2021axw}
W.-F. Chang, ``{One colorful resolution to the neutrino mass generation, three
  lepton flavor universality anomalies, and the Cabibbo angle anomaly},''
  \href{http://arxiv.org/abs/2105.06917}{{\ttfamily arXiv:2105.06917
  [hep-ph]}}.

\bibitem{Jueid:2021avn}
A.~Jueid, J.~Kim, S.~Lee, and J.~Song, ``{Type-X two Higgs doublet model in
  light of the muon $\mathbf{g-2}$: confronting Higgs and collider data},''
  \href{http://arxiv.org/abs/2104.10175}{{\ttfamily arXiv:2104.10175
  [hep-ph]}}.

\bibitem{Leveille:1977rc}
J.~P. Leveille, ``{The Second Order Weak Correction to (G-2) of the Muon in
  Arbitrary Gauge Models},''
\href{http://dx.doi.org/10.1016/0550-3213(78)90051-2}{{\em Nucl. Phys.}
  {\bfseries B137} (1978) 63--76}.

\bibitem{Aad:2014yka}
{\bfseries ATLAS} Collaboration, G.~Aad {\em et~al.}, ``{Search for the direct
  production of charginos, neutralinos and staus in final states with at least
  two hadronically decaying taus and missing transverse momentum in $pp$
  collisions at $\sqrt{s}$ = 8 TeV with the ATLAS detector},''
  \href{http://dx.doi.org/10.1007/JHEP10(2014)096}{{\em JHEP} {\bfseries 10}
  (2014) 096},
\href{http://arxiv.org/abs/1407.0350}{{\ttfamily arXiv:1407.0350 [hep-ex]}}.

\bibitem{Sirunyan:2018nwe}
{\bfseries CMS} Collaboration, A.~M. Sirunyan {\em et~al.}, ``{Search for
  supersymmetric partners of electrons and muons in proton-proton collisions at
  $\sqrt{s}=$ 13 TeV},''
  \href{http://dx.doi.org/10.1016/j.physletb.2019.01.005}{{\em Phys. Lett.}
  {\bfseries B790} (2019) 140--166},
\href{http://arxiv.org/abs/1806.05264}{{\ttfamily arXiv:1806.05264 [hep-ex]}}.

\bibitem{Sirunyan:2018vig}
{\bfseries CMS} Collaboration, A.~M. Sirunyan {\em et~al.}, ``{Search for
  supersymmetry in events with a $\tau$ lepton pair and missing transverse
  momentum in proton-proton collisions at $\sqrt{s} =$ 13 TeV},''
  \href{http://dx.doi.org/10.1007/JHEP11(2018)151}{{\em JHEP} {\bfseries 11}
  (2018) 151},
\href{http://arxiv.org/abs/1807.02048}{{\ttfamily arXiv:1807.02048 [hep-ex]}}.

\bibitem{LEP:2003aa}
{\bfseries LEP, ALEPH, DELPHI, L3, OPAL, LEP Electroweak Working Group, SLD
  Electroweak Group, SLD Heavy Flavor Group} Collaboration, t.~S. Electroweak,
  ``{A Combination of preliminary electroweak measurements and constraints on
  the standard model},''
\href{http://arxiv.org/abs/hep-ex/0312023}{{\ttfamily arXiv:hep-ex/0312023
  [hep-ex]}}.

\bibitem{ALEPH:2013dgf}
{\bfseries ALEPH, DELPHI, L3, OPAL, LEP Electroweak} Collaboration, S.~Schael
  {\em et~al.}, ``{Electroweak Measurements in Electron-Positron Collisions at
  W-Boson-Pair Energies at LEP},''
  \href{http://dx.doi.org/10.1016/j.physrep.2013.07.004}{{\em Phys. Rept.}
  {\bfseries 532} (2013) 119--244},
  \href{http://arxiv.org/abs/1302.3415}{{\ttfamily arXiv:1302.3415 [hep-ex]}}.

\bibitem{Carena:2004xs}
M.~Carena, A.~Daleo, B.~A. Dobrescu, and T.~M.~P. Tait, ``{$Z^\prime$ gauge
  bosons at the Tevatron},''
  \href{http://dx.doi.org/10.1103/PhysRevD.70.093009}{{\em Phys. Rev. D}
  {\bfseries 70} (2004) 093009},
  \href{http://arxiv.org/abs/hep-ph/0408098}{{\ttfamily arXiv:hep-ph/0408098}}.

\bibitem{Das:2021esm}
A.~Das, P.~S.~B. Dev, Y.~Hosotani, and S.~Mandal, ``{Probing the minimal
  $U(1)_X$ model at future electron-positron colliders via the fermion
  pair-production channel},'' \href{http://arxiv.org/abs/2104.10902}{{\ttfamily
  arXiv:2104.10902 [hep-ph]}}.

\bibitem{Peskin:1990zt}
M.~E. Peskin and T.~Takeuchi, ``{A New constraint on a strongly interacting
  Higgs sector},''
\href{http://dx.doi.org/10.1103/PhysRevLett.65.964}{{\em Phys. Rev. Lett.}
  {\bfseries 65} (1990) 964--967}.

\bibitem{Lavoura:1992np}
L.~Lavoura and J.~P. Silva, ``{The Oblique corrections from vector - like
  singlet and doublet quarks},''
\href{http://dx.doi.org/10.1103/PhysRevD.47.2046}{{\em Phys. Rev.} {\bfseries
  D47} (1993) 2046--2057}.

\bibitem{Chen:2017hak}
C.-Y. Chen, S.~Dawson, and E.~Furlan, ``{Vectorlike fermions and Higgs
  effective field theory revisited},''
  \href{http://dx.doi.org/10.1103/PhysRevD.96.015006}{{\em Phys. Rev.}
  {\bfseries D96} no.~1, (2017) 015006},
\href{http://arxiv.org/abs/1703.06134}{{\ttfamily arXiv:1703.06134 [hep-ph]}}.

\bibitem{Zyla:2020zbs}
{\bfseries Particle Data Group} Collaboration, P.~Zyla {\em et~al.}, ``{Review
  of Particle Physics},'' \href{http://dx.doi.org/10.1093/ptep/ptaa104}{{\em
  PTEP} {\bfseries 2020} no.~8, (2020) 083C01}.

\bibitem{Farzan:2012hh}
Y.~Farzan and A.~R. Akbarieh, ``{VDM: A model for Vector Dark Matter},''
  \href{http://dx.doi.org/10.1088/1475-7516/2012/10/026}{{\em JCAP} {\bfseries
  10} (2012) 026}, \href{http://arxiv.org/abs/1207.4272}{{\ttfamily
  arXiv:1207.4272 [hep-ph]}}.

\bibitem{Hisano:2010yh}
J.~Hisano, K.~Ishiwata, N.~Nagata, and M.~Yamanaka, ``{Direct Detection of
  Vector Dark Matter},'' \href{http://dx.doi.org/10.1143/PTP.126.435}{{\em
  Prog. Theor. Phys.} {\bfseries 126} (2011) 435--456},
  \href{http://arxiv.org/abs/1012.5455}{{\ttfamily arXiv:1012.5455 [hep-ph]}}.

\bibitem{Hoferichter:2017olk}
M.~Hoferichter, P.~Klos, J.~Men\'endez, and A.~Schwenk, ``{Improved limits for
  Higgs-portal dark matter from LHC searches},''
  \href{http://dx.doi.org/10.1103/PhysRevLett.119.181803}{{\em Phys. Rev.
  Lett.} {\bfseries 119} no.~18, (2017) 181803},
  \href{http://arxiv.org/abs/1708.02245}{{\ttfamily arXiv:1708.02245
  [hep-ph]}}.

\bibitem{XENON:2018voc}
{\bfseries XENON} Collaboration, E.~Aprile {\em et~al.}, ``{Dark Matter Search
  Results from a One Ton-Year Exposure of XENON1T},''
  \href{http://dx.doi.org/10.1103/PhysRevLett.121.111302}{{\em Phys. Rev.
  Lett.} {\bfseries 121} no.~11, (2018) 111302},
  \href{http://arxiv.org/abs/1805.12562}{{\ttfamily arXiv:1805.12562
  [astro-ph.CO]}}.

\bibitem{PandaX:2021osp}
{\bfseries PandaX} Collaboration, Y.~Meng {\em et~al.}, ``{Dark Matter Search
  Results from the PandaX-4T Commissioning Run},''
  \href{http://arxiv.org/abs/2107.13438}{{\ttfamily arXiv:2107.13438
  [hep-ex]}}.

\bibitem{Griest:1990kh}
K.~Griest and D.~Seckel, ``{Three exceptions in the calculation of relic
  abundances},'' \href{http://dx.doi.org/10.1103/PhysRevD.43.3191}{{\em Phys.
  Rev. D} {\bfseries 43} (1991) 3191--3203}.

\bibitem{Belanger:2020gnr}
G.~Belanger, A.~Mjallal, and A.~Pukhov, ``{Recasting direct detection limits
  within micrOMEGAs and implication for non-standard Dark Matter scenarios},''
  \href{http://dx.doi.org/10.1140/epjc/s10052-021-09012-z}{{\em Eur. Phys. J.
  C} {\bfseries 81} no.~3, (2021) 239},
  \href{http://arxiv.org/abs/2003.08621}{{\ttfamily arXiv:2003.08621
  [hep-ph]}}.

\bibitem{Alloul:2013bka}
A.~Alloul, N.~D. Christensen, C.~Degrande, C.~Duhr, and B.~Fuks, ``{FeynRules
  2.0 - A complete toolbox for tree-level phenomenology},''
  \href{http://dx.doi.org/10.1016/j.cpc.2014.04.012}{{\em Comput. Phys.
  Commun.} {\bfseries 185} (2014) 2250--2300},
  \href{http://arxiv.org/abs/1310.1921}{{\ttfamily arXiv:1310.1921 [hep-ph]}}.

\bibitem{Planck:2018vyg}
{\bfseries Planck} Collaboration, N.~Aghanim {\em et~al.}, ``{Planck 2018
  results. VI. Cosmological parameters},''
  \href{http://dx.doi.org/10.1051/0004-6361/201833910}{{\em Astron. Astrophys.}
  {\bfseries 641} (2020) A6}, \href{http://arxiv.org/abs/1807.06209}{{\ttfamily
  arXiv:1807.06209 [astro-ph.CO]}}.

\bibitem{Muoncollider}
K.~R. Long {\em et~al.}, ``{Muon colliders to expand frontiers of particle
  physics},'' \href{http://dx.doi.org/10.1038/s41567-020-01130-x}{{\em Nature
  Physics} {\bfseries 17} (2021) 289}.

\end{thebibliography}\endgroup
\end{document}